\documentclass[aps,prb,reprint,superscriptaddress,longbibliography]{revtex4-2}
\usepackage{amsmath,amssymb,amsthm,bm}
\usepackage{graphicx}
\usepackage{listings}
\usepackage[colorlinks=true,linkcolor=blue,citecolor=blue,urlcolor=blue]{hyperref}

\newtheorem{theorem}{Theorem}
\newtheorem{proposition}{Proposition}
\newtheorem{lemma}{Lemma}
\newtheorem{corollary}{Corollary}
\theoremstyle{definition}
\newtheorem{definition}{Definition}

\DeclareMathOperator{\Tr}{Tr}
\DeclareMathOperator{\diag}{diag}
\DeclareMathOperator{\SF}{SF}
\DeclareMathOperator{\Ind}{Ind}
\newcommand{\gref}{\gamma_{\mathrm{ref}}}
\newcommand{\cD}{\mathcal D}
\newcommand{\cV}{\mathcal V}

\begin{document}

\title{Interface invariants of rhombohedral graphene \\ with layer-dependent potentials and couplings}

\author{Yuelong Li}
\affiliation{Committee on Computational and Applied Mathematics, University of Chicago, Chicago, Illinois 60637, USA}

\date{\today}

\begin{abstract}
Domain walls in the perpendicular bias of rhombohedral (ABC-stacked) graphene bind valley-resolved interface modes whose signed number is fixed by a bulk-difference invariant. We determine which values of this invariant are reachable when the layer potentials are arbitrary monotone profiles, as produced by screening, and when the interlayer couplings vary from bond to bond. Within the single-particle continuum model, the invariant of an $m$-layer stack equals the difference of occupied angular weights of the zero-momentum levels, which reduces the classification to the ordering of those levels. At equal coupling, nonlinear monotone potentials reproduce exactly the set of invariants of linear potentials, also under the requirement of a common energy gap: screening moves phase boundaries but creates no new invariant values. With bond-dependent coupling, we characterize all occupied configurations. The range of attainable invariants is unchanged, but the gaps inside it are filled, beginning at five layers. Reflection-symmetric stacks realize a large subfamily with a guaranteed common gap, and lower bounds on single-bulk gaps complete the set of invariants with a common global gap for five and six layers. Among the examples is an eight-layer domain wall defined purely by gates, whose number of co-propagating channels per valley switches from 8 to 16 as the gate strength increases. Exact transfer-matrix calculations of the interface spectrum reproduce every predicted signed crossing count. A norm bound gives the minimum spread of interlayer coupling compatible with a prescribed gap.
\end{abstract}

\maketitle

\section{Introduction}
\label{sec:intro}

\begin{figure}
\includegraphics[width=\columnwidth]{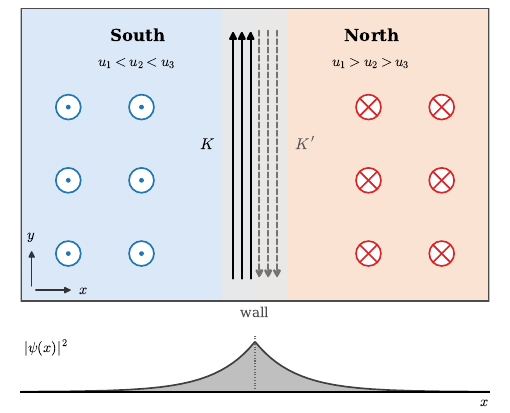}
\caption{Top view of a gate-defined domain wall. Split gates apply perpendicular fields of opposite sign on the two sides of $x=0$ ($\odot$ and $\otimes$), so the layer potentials are ordered oppositely in the South and North bulks. The interface channels run along the wall, in the $y$ direction. For the three-layer wall i of Table~\ref{tab:examples}, $\mathfrak C=3$: valley $K$ carries three channels along $+y$ (solid) and valley $K'$ three along $-y$ (dashed), for each spin. Bottom: the transverse probability density of a channel, localized at the wall and decaying into both bulks (schematic).}
\label{fig:topview}
\end{figure}

Multilayer graphene in a perpendicular electric field is a gate-controlled platform for one-dimensional topological channels. Their low-energy electrons occupy two inequivalent valleys, $K$ and $K'$, which time reversal exchanges. A perpendicular field applied by top and bottom gates opens a band gap and gives each valley a nonzero band invariant, with opposite signs in the two valleys~\cite{Martin,ZhangMM}. Along a line where the field reverses, each valley binds chiral modes, two per spin in a bilayer, and the modes of $K$ and $K'$ propagate in opposite directions~\cite{Martin,ZhangMM,Jung}. Summed over the two valleys the invariant vanishes exactly, so such a wall carries no net charge Hall current; its channels, the kink states, are valley helical. They survive because the valleys are far apart in momentum: a wall that varies slowly on the lattice scale cannot scatter an electron from $K$ to $K'$, so each valley conducts independently. Rhombohedral (ABC-stacked) multilayer graphene extends the construction to larger winding. Near each valley, $m$ layers produce chiral bands of winding $m$~\cite{MinMacDonald,ZhangABC}, and a field reversal binds $m$ co-propagating channels per valley and spin~\cite{Jung}. Figure~\ref{fig:topview} shows such a wall from above.

The measured quantity is the conductance along the wall. Without intervalley scattering each channel carries $e^2/h$, and an ideal bilayer wall, with four channels in each direction, has the two-terminal conductance $4e^2/h$. Ju \emph{et al.}~\cite{Ju} located AB--BA stacking domain walls in bilayer graphene by near-field infrared imaging and found one-dimensional conducting channels along them, with a ballistic length of about 400~nm at 4~K. Li \emph{et al.}~\cite{LiNN} defined the wall electrostatically, with aligned split top and bottom gates that apply opposite fields on the two sides of a narrow line; this is the geometry of Figs.~\ref{fig:topview} and~\ref{fig:setup}(a) and of the model studied here. Later devices of this type operated as a valley valve and beam splitter~\cite{Li2018} and showed resistance plateaus quantized to $h/4e^2$ at zero magnetic field, persisting to tens of kelvin~\cite{Huang2024}. The bulk valley Hall response has been detected separately through nonlocal transport~\cite{Gorbachev,Sui,Shimazaki}. In rhombohedral stacks, interactions in the flat low-energy bands can polarize the valleys spontaneously~\cite{ZhangSQH}. Han \emph{et al.}~\cite{Han} observed an integer quantum anomalous Hall effect with Chern number $|C|=5$ in rhombohedral pentalayer graphene at zero magnetic field, establishing that a five-layer stack can host an integer state of this size; fractional states appear in a moir\'e pentalayer~\cite{Lu}. These measurements probe the outer edge of a sample, where the stack meets vacuum. The invariant classified here describes a different boundary: an interface between two gated bulks of the same stack.

Two features of this platform make it mathematically tractable. First, a straight wall is translation invariant along its length $y$. The longitudinal momentum $k_y$ is then conserved, and one valley of the device is described by a one-parameter family of self-adjoint Dirac-type operators acting in the transverse coordinate $x$,
\begin{equation}
H_I(k_y)=-i\Sigma_1\partial_x+k_y\Sigma_2+V(x),
\label{eq:interface}
\end{equation}
on $H^1(\mathbb R;\mathbb C^{2m})$. Here $\Sigma_\alpha=I_m\otimes\sigma_\alpha$ acts on the sublattice index of each layer, $V(x)$ contains the layer potentials and interlayer couplings at position $x$ [Eq.~\eqref{eq:model} below], and the bulks are South at $x\to-\infty$ and North at $x\to+\infty$. The spectrum of $H_I(k_y)$ consists of the two bulk continua and, inside their common gap, discrete interface eigenvalues $E_n(k_y)$. A channel at the Fermi energy $E_F$ is a crossing $E_n(k_y)=E_F$, and it propagates along $+y$ or $-y$ according to the sign of $dE_n/dk_y$. The signed number of crossings as $k_y$ runs over $\mathbb R$ is the spectral flow of $H_I(k_y)-E_F$. It cannot change under deformations of the wall that keep $E_F$ inside the gap of both bulks, because branches can then neither appear nor disappear at $E_F$. Second, a single valley generically has no antiunitary symmetry, since time reversal maps it to the other valley. In the Altland--Zirnbauer classification it therefore belongs to class A, whose two-dimensional invariant is an integer of Chern type; the time-reversal symmetric pair of valleys has total invariant zero.

The spectral flow is also what a transport measurement sees. Let $H_I=-i\Sigma_1\partial_x-i\Sigma_2\partial_y+V(x)$ be the two-dimensional operator whose Fourier components along the wall are the $H_I(k_y)$, and choose two smooth switches: $\chi(y)$, increasing from 0 to 1 across a line $y=y_0$ that cuts the wall, and $\varphi(E)$, increasing from 0 to 1 inside the common bulk gap. The interface conductivity is~\cite{BalInterface,QuinnBal}
\begin{equation}
\sigma_I=\Tr\bigl(i[H_I,\chi(y)]\,\varphi'(H_I)\bigr).
\label{eq:intro-sigma}
\end{equation}
Here $i[H_I,\chi]=\Sigma_2\chi'(y)$ is the current through the line $y=y_0$, and $\varphi'$ is a unit-weight energy window inside the gap, the analog of a small bias voltage. Translation invariance reduces the trace to Fourier modes with measure $dk_y/2\pi$, and only interface states contribute because $\varphi'$ vanishes on the bulk continua. Since $\Sigma_2=\partial H_I/\partial k_y$, the Hellmann--Feynman identity gives $\langle\psi_n|\Sigma_2|\psi_n\rangle=dE_n/dk_y$, and each term becomes a total derivative:
\begin{align}
2\pi\sigma_I&=\sum_n\int_{\mathbb R}\frac{d}{dk_y}\,
\varphi\bigl(E_n(k_y)\bigr)\,dk_y\notag\\
&=\SF_{k_y\uparrow}\bigl(H_I(k_y)-E_F\bigr)
\label{eq:intro-sigma-sf}
\end{align}
for any $E_F$ inside the window. A branch that crosses the window upward contributes $+1$ and one that crosses it downward contributes $-1$, whatever the choice of $\chi$ and $\varphi$. In units with $e=\hbar=1$ a single channel has conductance $e^2/h=1/2\pi$, so $2\pi\sigma_I$ is the conductance of the wall in units of $e^2/h$, for one valley and one spin.

The bulk--interface correspondence ties the spectral flow to the two bulks. Because the continuum Hamiltonian is posed on the whole momentum plane, the Chern integral $I_B$ of a single bulk need not be an integer: for $m$ layers it lies in $\mathbb Z-m^2/2$, which includes the familiar valley Chern numbers $\pm m/2$. The difference between two bulks with the same high-momentum behavior is an integer, the bulk-difference invariant $\mathfrak C=I_N-I_S$, and under a common energy gap (Sec.~\ref{sec:model}B) it equals the spectral flow~\cite{BalInterface,FrazierBal},
\begin{equation}
\SF_{k_y\uparrow}\bigl(H_I(k_y)-E_F\bigr)=\mathfrak C=I_N-I_S.
\label{eq:intro-flow}
\end{equation}
(BDI here abbreviates this invariant, not the symmetry class of the same name.) Gates change $\mathfrak C$ only where a bulk gap closes, so predicting the channel counts a device can show amounts to classifying the values of $\mathfrak C$ that the allowed bulk parameters reach.

Frazier and Bal~\cite{FrazierBal} carried out this classification for gated ABC graphene in the continuum model under two idealizations: a field that is uniform through the stack, so that the layer potentials are linear in the layer index, and a single interlayer coupling common to all bonds. They name non-uniform fields and layer-dependent couplings as the natural next steps~\cite{FrazierBal}. Both idealizations can fail in devices. Interlayer screening makes the layer potentials nonlinear in the layer index~\cite{Koshino}, and staged intercalation or other gallery-selective modifications change individual interlayer bonds~\cite{Dresselhaus,Ahmad}. We ask which invariant values each source of layer dependence can reach. The question separates two kinds of control: moving the boundaries between known phases, and creating invariant values that no uniform configuration attains.

Our answers are as follows. The BDI equals the difference of the occupied angular weights of the zero-momentum levels (Theorem~\ref{thm:weights}), so classification reduces to deciding which zero-momentum levels can lie below the gap. At equal coupling, arbitrary monotone potentials give exactly the set of invariants of linear potentials, including when both bulks must share an energy gap (Theorem~\ref{thm:equal}). With bond-dependent coupling, we characterize every occupied configuration (Theorem~\ref{thm:configurations}). The range of the set of invariants is unchanged, but unequal coupling fills the gaps inside it; the first new values appear at $m=5$. A reflection-symmetric subfamily comes with a common energy gap by construction (Theorem~\ref{thm:symmetric}), and lower bounds on single-bulk gaps complete the set of invariants with a common global gap at five and six layers. We illustrate the results with examples and verify every predicted count against exact transfer-matrix calculations of the interface spectrum. The most experimentally natural example is an eight-layer stack with a fixed bond pattern, in which increasing the gate strength switches the wall from 8 to 16 co-propagating channels per valley. Finally, we bound from below the coupling spread required for a new invariant with a prescribed gap.

Throughout, the invariant refers to one valley and one spin sector of a single-particle model, and its transport interpretation assumes negligible intervalley scattering. Section~\ref{sec:model} sets up the model and the weight formula. Sections~\ref{sec:equal} and~\ref{sec:unequal} treat equal and bond-dependent coupling, Sec.~\ref{sec:numerics} presents the interface spectra, and Sec.~\ref{sec:discussion} discusses physical scales and limitations. Proofs and numerical details are in the appendices.

\section{Model and invariant}
\label{sec:model}

\subsection{Continuum model}

We use the $\pi$-orbital continuum model of an ABC stack with $m\ge2$ layers~\cite{MinMacDonald,FrazierBal}, retaining the intralayer Dirac term and vertical nearest-layer coupling. In valley $K_\tau$ and one spin sector, the envelope spinor is $\Psi=(\psi_{A_1},\psi_{B_1},\ldots,\psi_{A_m},\psi_{B_m})^{\mathsf T}$, and the intralayer term is $\hbar v(k_x\sigma_1+\tau k_y\sigma_2)$ with $\bm k$ measured from $K_\tau$. We set $\hbar=v=1$ and $\tau=+1$; the other valley follows from $k_y\mapsto-k_y$ and reverses every invariant below. In the chosen registry the $j$th interlayer bond joins $B_j$ to $A_{j+1}$.

A bulk $B\in\{N,S\}$ is specified by layer potentials $\bm u^B=(u^B_1,\ldots,u^B_m)$, acting equally on both sublattices of a layer, and positive couplings $\bm\gamma^B=(\gamma^B_1,\ldots,\gamma^B_{m-1})$. Its Bloch Hamiltonian is
\begin{align}
H_B(\bm k)&=\bigoplus_{j=1}^m\bigl(u_j^BI_2+k_x\sigma_1+k_y\sigma_2\bigr)
\notag\\
&\quad+\sum_{j=1}^{m-1}\gamma^B_j\bigl(|B_j\rangle\langle A_{j+1}|
+|A_{j+1}\rangle\langle B_j|\bigr).
\label{eq:model}
\end{align}
We write $V_B:=H_B(\bm 0)$. Skew and remote interlayer couplings, spin--orbit coupling, intervalley mixing, and interactions are omitted, and the potentials and couplings are prescribed inputs rather than solutions of a self-consistent problem.

\begin{figure}
\includegraphics[width=\columnwidth]{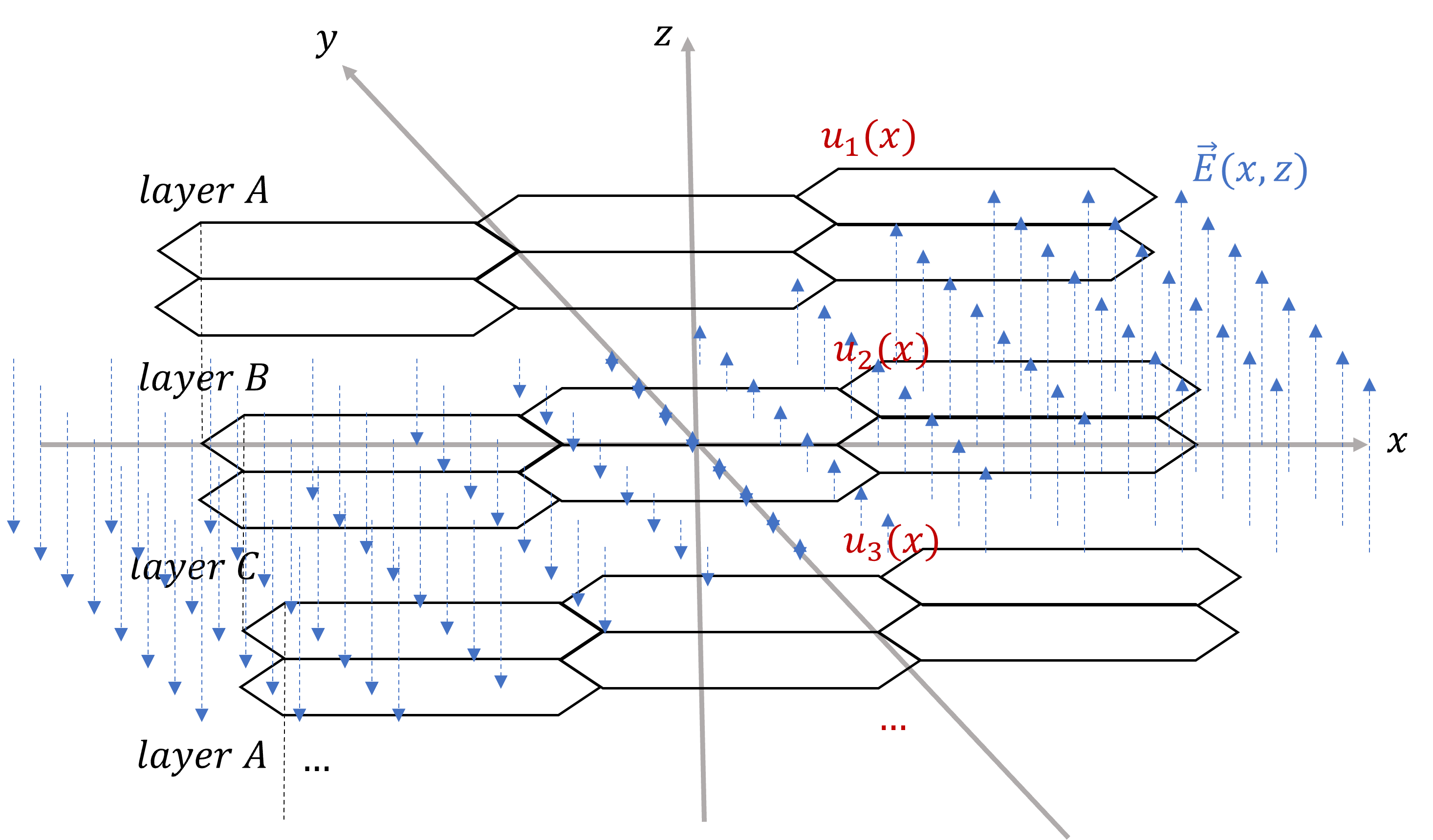}\\[2pt]
\includegraphics[width=0.62\columnwidth]{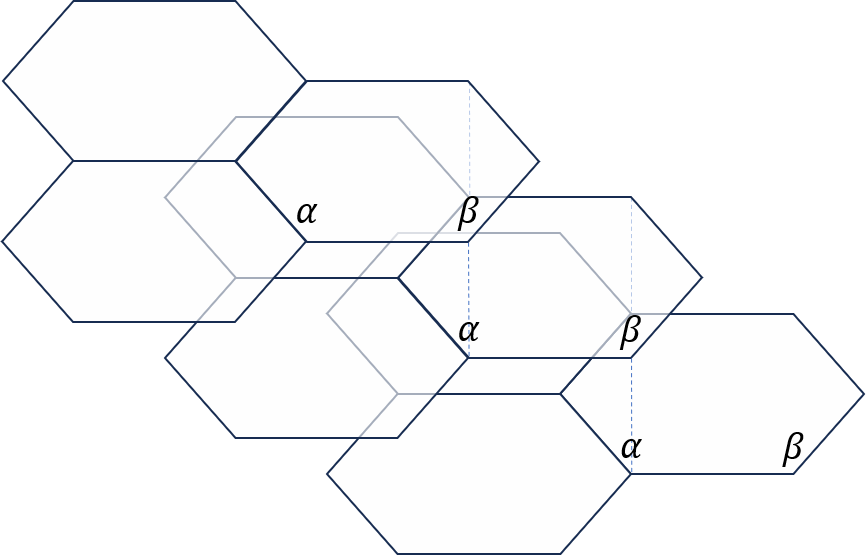}
\caption{(a) Geometry of the domain wall. ABC-stacked layers lie parallel to the $xy$ plane; here $A$, $B$, and $C$ label stacking registries, not sublattices. A perpendicular field $\bm E(x,z)$ reverses sign across $x=0$, producing layer potentials $u_j(x)$ that depend only on $x$ and the layer index. (b) ABC stacking in the chosen sublattice convention, with $\alpha\equiv A$ and $\beta\equiv B$. The vertical interlayer bonds (dashed) join $\beta$ of layer $j$ to $\alpha$ of layer $j+1$, that is, $B_j$ to $A_{j+1}$.}
\label{fig:setup}
\end{figure}

Each bulk is \emph{monotone}: its layer potentials either never decrease or never increase with the layer index. We record which by the \emph{orientation} $\epsilon_B\in\{+1,-1\}$, defined by $\epsilon_B(u^B_{j+1}-u^B_j)\ge0$ for all $j$; thus $\epsilon_B=+1$ if the potentials increase through the stack and $\epsilon_B=-1$ if they decrease. Physically, $\epsilon_B$ is the sign of the perpendicular field in bulk $B$, up to the convention of which layer is on top. A field-reversal wall, as in Fig.~\ref{fig:topview}, has $\epsilon_N=-\epsilon_S$; a wall with $\epsilon_N=\epsilon_S$ changes only the strength of the field. We restrict to monotone bulks; the weight formula below holds without this assumption. Figure~\ref{fig:setup} shows the geometry and the bond convention.

\subsection{The bulk invariant}

All our computations rest on one quantity, the curvature integral $I_B$ of a single bulk. Equation~\eqref{eq:intro-flow}, through which the conductivity and the spectral flow enter, only serves to interpret $\mathfrak C=I_N-I_S$.

Let $e^B_1(r)\le\cdots\le e^B_{2m}(r)$ be the eigenvalues of $H_B$ at momentum $|\bm k|=r$; they depend only on $r$ [Eq.~\eqref{eq:covariance} below]. A bulk has a \emph{direct gap} if $\delta_B:=\inf_{r\ge0}[e^B_{m+1}(r)-e^B_m(r)]>0$. For positive couplings this is equivalent to its zero-momentum version, $e^B_m(0)<e^B_{m+1}(0)$ (Lemma~\ref{lem:zero-gap}). The projector $P_B(\bm k)$ onto the lowest $m$ bands is then smooth, and we set
\begin{equation}
I_B=\frac{i}{2\pi}\int_{\mathbb R^2}
\Tr\bigl(P_B[\partial_{k_x}P_B,\partial_{k_y}P_B]\bigr)\,d^2k.
\label{eq:curvature}
\end{equation}
The individual integrals need not be integers on the noncompact momentum plane. The two projectors share their directional limit at infinity (Appendix~\ref{app:weights}), so $\mathfrak C:=I_N-I_S$ is an integer. A direct gap in each bulk is all that $\mathfrak C$ requires.

\begin{figure}
\includegraphics[width=\columnwidth]{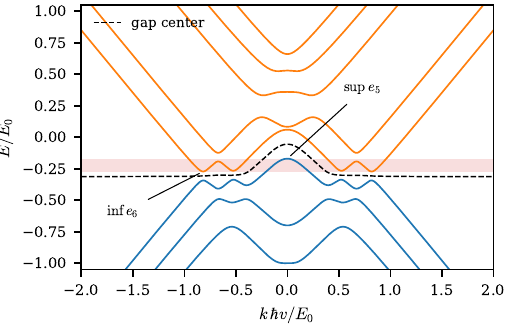}
\caption{Radial spectrum of a five-layer bulk with $\bm u=(-1,-1,0,0.36,0.36)E_0$ and $\gamma_j=0.3E_0$ (lower bands blue, upper bands orange). The direct gap stays open at every momentum (minimum $0.071E_0$), but $\sup_k e_5=-0.170E_0$, reached at $k=0$, lies above $\inf_k e_6=-0.272E_0$, reached near $|k|\simeq0.8E_0/\hbar v$. No fixed energy interval is avoided, so the bulk has no global gap (shaded overlap). The dashed curve is the gap center $(e_5+e_6)/2$.}
\label{fig:directgap}
\end{figure}

\emph{Common global gap.} Reading $\mathfrak C$ as a channel count requires more: the switch $\varphi$ in Eq.~\eqref{eq:intro-sigma} must rise inside an energy interval that neither bulk spectrum enters at any momentum. The pair has a \emph{common global gap} if
\begin{equation}
E_{\mathrm v}:=\max_B\sup_r e^B_m(r)<
E_{\mathrm c}:=\min_B\inf_r e^B_{m+1}(r),
\label{eq:global-gap}
\end{equation}
and we write $G=E_{\mathrm c}-E_{\mathrm v}$. A single bulk has a \emph{global gap} if $\sup_r e^B_m<\inf_r e^B_{m+1}$. A direct gap alone need not leave any fixed energy interval empty; Fig.~\ref{fig:directgap} shows a bulk with a direct gap but no global gap.

With such a gap containing $E_F$, and wall coefficients that are smooth, bounded, and constant outside a strip, Eq.~\eqref{eq:intro-flow} holds~\cite{BalInterface,QuinnBal,FrazierBal}; its sign, with crossings of positive slope counted as $+1$, is confirmed numerically in Sec.~\ref{sec:numerics}.

\emph{Sets of invariants.} Fix $m$ and the orientations $\epsilon_N$ and $\epsilon_S$ of the two bulks (Sec.~\ref{sec:model}A), and write $\mathcal O=(\epsilon_N,\epsilon_S)$. For each class of bulk pairs below, with potential offsets and slopes otherwise unrestricted and chosen independently in the two bulks, we collect the values of $\mathfrak C$:
\begin{itemize}\setlength{\itemsep}{1pt}\setlength{\parskip}{0pt}
\item $\cV_{\mathrm{dir}}(m;\mathcal O)$: monotone potentials, positive couplings, and a direct gap in each bulk;
\item $\cV_{\mathrm{glob}}(m;\mathcal O)$: the same, with a common global gap;
\item $\cV_{\mathrm{eq}}(m;\mathcal O)$: equal couplings, $\gamma^B_j=\gref$ for all $j$ and $B$ and a fixed $\gref>0$, with a direct gap in each bulk;
\item $\cV_{\mathrm{sym}}(m;\mathcal O)$: reflection-symmetric bulks, $u_{m+1-j}=-u_j$ and $\gamma_{m-j}=\gamma_j$, with a direct gap in each.
\end{itemize}
By Eq.~\eqref{eq:intro-flow}, $\cV_{\mathrm{glob}}$ is the set of signed channel counts that walls between such bulks can carry. The set $\cV_{\mathrm{dir}}$ records band-projector invariants only; it contains $\cV_{\mathrm{glob}}$ and is easier to compute. Sections~\ref{sec:equal} and~\ref{sec:unequal} compute all four sets and show that they are nested,
\begin{equation}
\cV_{\mathrm{eq}}\subseteq\cV_{\mathrm{sym}}\subseteq\cV_{\mathrm{glob}}
\subseteq\cV_{\mathrm{dir}},
\label{eq:nesting}
\end{equation}
with the values for $5\le m\le8$ listed in Table~\ref{tab:invariants}.

\subsection{Occupied angular weights}

\begin{figure}
\includegraphics[width=\columnwidth]{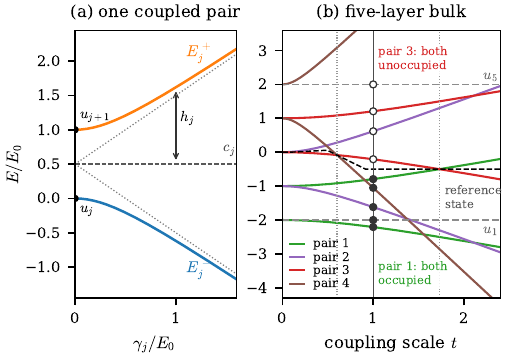}
\caption{(a) The two levels of one coupled pair, Eq.~\eqref{eq:pair}, against its coupling, for $u_j=0$ and $u_{j+1}=E_0$. They start at $u_j$ and $u_{j+1}$, lie symmetrically about the center $c_j$ at distance $h_j$, and approach $c_j\pm\gamma_j$ (dotted). (b) All zero-momentum levels of a five-layer bulk with linear potentials $\bm u=(-2,-1,0,1,2)E_0$ and couplings $t\,(0.5,1,0.5,2.5)E_0$. Colors label the coupled pairs, gray dashed lines the unpaired levels $u_1$ and $u_5$, and the black dashed line the center of the zero-momentum gap. For $t>\sqrt3$ every pair straddles the gap (the reference state). At $t=\sqrt3$ the lower level of the central pair 3 leaves the occupied levels and the upper level of pair 1 enters; at $t\simeq0.60$ the lower level of pair 4 replaces that of pair 3 (dotted lines). At $t=1$ the five lowest levels are occupied (filled circles): both levels of pair 1 lie below the gap and both levels of pair 3 above it, while pairs 2 and 4 straddle it.}
\label{fig:pairs}
\end{figure}

With the diagonal generator $D=\bigoplus_{j=1}^m\diag(j-1,j)$, whose entries agree on the two ends of every interlayer bond, the symbol is rotationally covariant:
\begin{equation}
H_B\bigl(r\cos\theta,r\sin\theta\bigr)
=e^{i\theta D}H_B(r,0)e^{-i\theta D}.
\label{eq:covariance}
\end{equation}
At zero momentum $V_B$ decomposes into the unpaired orbitals $A_1$ and $B_m$, with energies $u_1$ and $u_m$, and the \emph{coupled pairs} $(B_j,A_{j+1})$ joined by bond $j$ (the dimer sites of Ref.~\cite{McCannKoshino}). The coupling term $\gamma_j(|B_j\rangle\langle A_{j+1}|+\text{h.c.})$ splits pair $j$ into the levels
\begin{equation}
E^\pm_j=\frac{u_j+u_{j+1}}2\pm
\sqrt{\frac{(u_{j+1}-u_j)^2}4+\gamma_j^2}
\label{eq:pair}
\end{equation}
for $1\le j\le m-1$. The two levels sit symmetrically about the center $c_j=(u_j+u_{j+1})/2$, at the half-splitting $h_j=\sqrt{(u_{j+1}-u_j)^2/4+\gamma_j^2}$. At $\gamma_j=0$ they start from $u_j$ and $u_{j+1}$, and for large $\gamma_j$ they approach $c_j\pm\gamma_j$ [Fig.~\ref{fig:pairs}(a)]. Each pair has its own center and splitting, so the levels of different pairs cross as the couplings change [Fig.~\ref{fig:pairs}(b)]. Here and below we suppress the bulk label when only one bulk is considered.

The generator acts as $0$ on $A_1$, as $m$ on $B_m$, and as $jI_2$ on pair $j$, so every zero-momentum level carries a parameter-independent \emph{angular weight}: $0$ for $u_1$, $j$ for $E^\pm_j$, and $m$ for $u_m$.

The curvature integral reduces to these weights in two steps. First, covariance turns the integrand into a total radial derivative. With $P(\theta,r)$ the projector $P_B$ in polar coordinates and $W_B(r)=\Tr\bigl(D\,P(0,r)\bigr)$, which equals $\Tr(D\,P(\theta,r))$ for every $\theta$,
\begin{equation}
i\Tr\bigl(P[\partial_rP,\partial_\theta P]\bigr)=-W_B'(r).
\label{eq:bridge}
\end{equation}
This identity involves no choice of eigenvector phases (Appendix~\ref{app:weights}). Since $d^2k=r\,dr\,d\theta$ and $[\partial_{k_x}P,\partial_{k_y}P] =r^{-1}[\partial_rP,\partial_\theta P]$, integrating it gives
\begin{equation}
I_B=W_B(0)-W_B(\infty).
\label{eq:boundary}
\end{equation}
Second, the two ends are evaluated directly. At large momentum every bulk approaches the same projector, with $W_B(\infty)=m^2/2$. At zero momentum the Hamiltonian is block diagonal in the unpaired orbitals and the coupled pairs, so $W_B(0)$ is the sum of the angular weights of the occupied levels.

\begin{theorem}[Occupied weights]
\label{thm:weights}
For positive couplings and a direct gap in each bulk,
\begin{equation}
I_B=W_B-\frac{m^2}{2},\qquad
\mathfrak C=W_N-W_S,
\label{eq:weight-formula}
\end{equation}
where $W_B:=W_B(0)$ is the sum of the angular weights of the $m$ lowest zero-momentum levels of bulk $B$.
\end{theorem}

The weight at infinity cancels in $\mathfrak C$, so $\mathfrak C$ changes only when a level crosses the middle gap at zero momentum. An exchange in bulk $B$ that replaces an occupied level of weight $w_{\mathrm{out}}$ by one of weight $w_{\mathrm{in}}$ changes $\mathfrak C$ by $\eta_B(w_{\mathrm{in}}-w_{\mathrm{out}})$, with $\eta_N=+1$ and $\eta_S=-1$. For weak linear bias, Eq.~\eqref{eq:weight-formula} gives $\mathfrak C=\pm m$, the familiar count of $m$ kink channels per valley~\cite{Martin,Jung}.

\section{Equal interlayer coupling}
\label{sec:equal}

Let every bond in both bulks carry the same coupling $\gref>0$. Here and in Sec.~\ref{sec:unequal} we fix the orientations of the two bulks, $\mathcal O=(\epsilon_N,\epsilon_S)$, because a bulk's orientation fixes the sign of its curvature [Eq.~\eqref{eq:oriented} below]. Where a single orientation is needed we use the field reversal of Fig.~\ref{fig:topview}, $\mathcal O=(-1,+1)$: North descending, South ascending. Put
\begin{equation}
p_{\max}=\Bigl\lfloor\frac m2\Bigr\rfloor-1,\qquad
f_m(p)=p\,(m-1-p).
\label{eq:counts}
\end{equation}
For a nondecreasing profile, equal coupling orders the upper levels among themselves and the lower levels among themselves: $E^\pm_1\le\cdots\le E^\pm_{m-1}$ for every $\gref>0$ (Lemma~\ref{lem:order}). The occupied lower levels and the occupied upper levels therefore each form a prefix, $\{E^-_1,\ldots,E^-_q\}$ and $\{E^+_1,\ldots,E^+_p\}$ with $p+q=m-1$, and the inequality $E^-_{q+1}<u_{q+1}\le u_{p+1}<E^+_p$ forces $p<q$, that is, $0\le p\le p_{\max}$ (Lemma~\ref{lem:prefix}). The occupied weight is $W_\uparrow=\binom m2-f_m(p)$. Reversing the layer order maps each weight $w$ to $m-w$, so for either orientation
\begin{equation}
W_\epsilon=\frac{m^2}{2}-\epsilon\Bigl[\frac m2+\Delta\Bigr],
\label{eq:oriented}
\end{equation}
with $\Delta=f_m(p)$ at equal coupling. Centered linear profiles realize every $p$ at fixed $\gref$ (Lemma~\ref{lem:linear}).

\begin{theorem}[Equal-coupling invariants]
\label{thm:equal}
For fixed $m$, $\gref$, and $\mathcal O$, with unrestricted independent potential offsets and slopes in the two bulks,
\begin{equation}
\cV_{\mathrm{eq}}(m;\mathcal O)=\Bigl\{\epsilon_S\Bigl[\tfrac m2+f_m(p_S)\Bigr]
-\epsilon_N\Bigl[\tfrac m2+f_m(p_N)\Bigr]\Bigr\},
\label{eq:equal-invariants}
\end{equation}
where $0\le p_N,p_S\le p_{\max}$. The same set is obtained with linear potentials, and with the additional requirement of a common global gap of any positive width.
\end{theorem}

\begin{figure}
\includegraphics[width=\columnwidth]{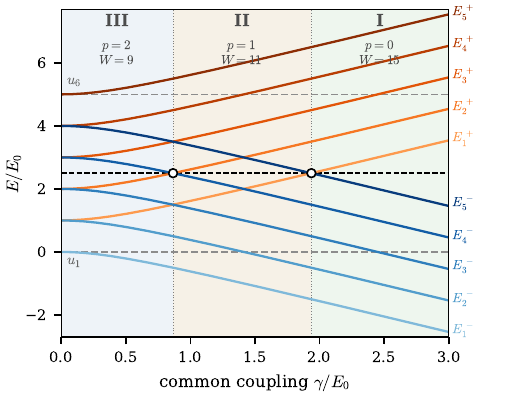}
\caption{Zero-momentum levels of the six-layer ascending profile $\bm u=(0,1,\ldots,5)E_0$ against the common coupling $\gamma$. Orange: upper levels $E^+_j$; blue: lower levels $E^-_j$; gray dashed: the unpaired levels $u_1$ and $u_6$. Levels within each family never cross (Lemma~\ref{lem:order}). By the reflection symmetry of the linear profile the gap center stays at $2.5E_0$ (black dashed), and the six levels below it are occupied. The gap closes where an upper level meets a lower level on that line (circles): at $\gamma=\sqrt{15}/2\simeq1.94$ between $E^+_1$ and $E^-_5$, and at $\gamma=\sqrt3/2\simeq0.87$ between $E^+_2$ and $E^-_4$. These closings separate phases I, II, and III, with $p=0,1,2$ occupied upper levels and occupied weights $W=15$, $11$, and $9$.}
\label{fig:branch-uniform}
\end{figure}

The statement about a common global gap uses the reflection symmetry of centered linear profiles: with $R$ reversing the component order and $Z=\diag(1,-1,\ldots)$, the unitary $J=ZR$ anticommutes with the radial symbol whenever the potentials are antisymmetric and the couplings palindromic (Lemma~\ref{lem:J}). The spectrum is then symmetric about zero, and the direct gap becomes a global gap around zero, so any two such bulks share a gap. For descending North and ascending South potentials the set of invariants is $\{m+f_m(p_N)+f_m(p_S)\}$. For $m=6$ it is $\{6,10,12,14,16,18\}$, and the symmetric walls $p_N=p_S$ reproduce the values of Ref.~\cite{FrazierBal}. Figure~\ref{fig:branch-uniform} shows the ordered families and the resulting phases for $m=6$.

Theorem~\ref{thm:equal} concerns which values are attainable. Screening moves the transitions between them but cannot create new ones. We do not claim that a given nonlinear profile can be deformed to its linear representative without closing the gap, or that each value is reachable under fixed bounds on fields or offsets.

\section{Bond-dependent coupling}
\label{sec:unequal}

\subsection{All occupied configurations}

Without equal coupling, the upper (or lower) levels need not be ordered among themselves, and the prefix structure fails. What survives is the inequality $E^-_q<u_q\le u_{p+1}<E^+_p$ for $q\le p+1$, which holds for any positive couplings and nondecreasing potentials. It turns out to be the only constraint.

\begin{definition}
A pair $(U,C)$ of subsets of $\{1,\ldots,m-1\}$ is \emph{admissible} if $|U|=|C|$ and $\max U+2\le\min C$ (vacuous when both are empty). For a bulk, $U$ collects the \emph{occupied pairs}, whose two levels are both occupied, and $C$ the \emph{unoccupied pairs}, whose two levels are both unoccupied; every other pair is a \emph{straddling pair}, with one level on each side of the gap [Fig.~\ref{fig:pairs}(b)]. The \emph{reference state} is $U=C=\varnothing$. Put $\Delta(U,C)=\sum_{j\in C}j-\sum_{j\in U}j$, and let $\cD_m$ be the set of values of $\Delta$ over admissible pairs. The number $\Delta$ measures how far a bulk's curvature lies from its value in the reference state: by Theorem~\ref{thm:configurations} and Eq.~\eqref{eq:oriented}, a bulk with orientation $\epsilon_B$ has $I_B=-\epsilon_B\bigl(\tfrac m2+\Delta\bigr)$, so the reference state gives $I_B=\mp\tfrac m2$, and each occupied or unoccupied pair moves $I_B$ further from zero. For a wall, $\mathfrak C=\epsilon_S(\tfrac m2+\Delta_S)-\epsilon_N(\tfrac m2+\Delta_N)$, which for the field reversal $\mathcal O=(-1,+1)$ reads $\mathfrak C=m+\Delta_N+\Delta_S$.
\end{definition}

\begin{theorem}[Occupied configurations]
\label{thm:configurations}
Let $\bm u$ be nondecreasing, $\bm\gamma>0$, and suppose the bulk has a direct gap. Then its occupied zero-momentum levels are
\begin{equation}
\{u_1\}\cup\{E^-_j:j\notin C\}\cup\{E^+_j:j\in U\}
\label{eq:configuration}
\end{equation}
for an admissible pair $(U,C)$, and $W_\uparrow=\binom m2-\Delta(U,C)$. Conversely, every admissible pair is realized by some strictly increasing $\bm u$ and positive $\bm\gamma$ with a direct gap.
\end{theorem}

\begin{proof}
If $u_1=u_m$, all potentials coincide, and the levels $u_1$ (twice) sit between $m-1$ lower and $m-1$ upper pair levels, so the middle gap closes. Hence $u_1<u_m$. Every upper level satisfies $E^+_j>u_{j+1}\ge u_1$, so at most $m-1$ levels lie below $u_1$, and the gap forces $u_1$ to be occupied. All $m-1$ lower levels and $u_1$ lie strictly below $u_m$, so $u_m$ is unoccupied. Let $U$ index the occupied upper levels and $C$ the unoccupied lower levels; since $E^-_j<E^+_j$, these are the occupied and unoccupied pairs. Counting occupied levels gives $|U|=|C|$. If $p\in U$ and $q\in C$ with $q\le p+1$, then $E^-_q<u_q\le u_{p+1}<E^+_p$ places an unoccupied level below an occupied one, which is impossible; hence $\max U+2\le\min C$.

Conversely, the pair $U=C=\varnothing$ is realized by Lemma~\ref{lem:linear} with $p=0$. Otherwise let $P=\max U$ and $K=\min C$, and fix $a>0$. Set $u_i=-a$ for $i\le P+1$, $u_i=a$ for $i\ge K$, and interpolate increasingly in between. Put $\gamma_j=a/2$ on $U\cup C$ and $\gamma_j=2a$ elsewhere. A pair $j\in U$ has $u_j=u_{j+1}=-a$ and both levels negative, and a pair in $C$ has both levels positive. Every other pair has $E^+_jE^-_j=u_ju_{j+1}-\gamma_j^2<0$, so it straddles zero. Exactly $1+2|U|+(m-1-|U|-|C|)=m$ levels are negative and none vanishes, so Lemma~\ref{lem:zero-gap} gives a direct gap. A small increasing perturbation makes $\bm u$ strictly monotone without changing the signs.
\end{proof}

\begin{figure}
\includegraphics[width=0.8\columnwidth]{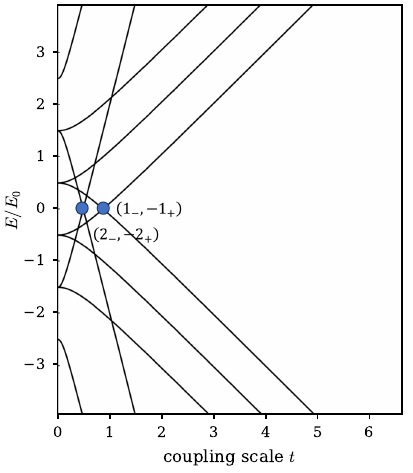}
\caption{Zero-momentum levels of a six-layer bulk with $\bm u=(-2.5,-1.5,\ldots,2.5)E_0$ and couplings $\gamma_j=t\,(4,1,1,1,4)E_0$, against the coupling scale $t$. The label $l_\pm$ denotes $E^\pm_{l+3}$, and circles mark exchanges across the middle gap, both at zero energy by reflection symmetry: $(E^-_5,E^+_1)$ at $t=\sqrt{15}/8\simeq0.484$ and $(E^-_4,E^+_2)$ at $t=\sqrt3/2\simeq0.866$. Between them the bulk has $U=\{2\}$ and $C=\{4\}$, so $\Delta=2$. Equal coupling forbids this configuration, because there the outer pair exchanges first (Fig.~\ref{fig:branch-uniform}); the strong outer bonds here delay that exchange until after the inner one.}
\label{fig:branch-unequal}
\end{figure}

For nonincreasing profiles, layer reversal again gives Eq.~\eqref{eq:oriented}, now with $\Delta\in\cD_m$. Equal coupling corresponds to the prefix pairs $U=\{1,\ldots,p\}$ and $C=\{m-p,\ldots,m-1\}$, for which $\Delta=f_m(p)$. Figure~\ref{fig:branch-unequal} shows how bond-dependent coupling reorders the exchanges and produces a non-prefix configuration.

\begin{corollary}[Range]
\label{cor:range}
$\cD_m\subseteq\{0\}\cup\{2,3,\ldots,f_m(p_{\max})\}$, and both endpoints are attained by prefix pairs. Consequently $\cV_{\mathrm{dir}}(m;\mathcal O)$ is obtained from Eq.~\eqref{eq:equal-invariants} by letting $f_m(p_B)$ range over $\cD_m$, and it has the same minimum and maximum as $\cV_{\mathrm{eq}}(m;\mathcal O)$.
\end{corollary}

\begin{proof}
Write $U=\{a_1<\dots<a_k\}$ and $C=\{b_1<\dots<b_k\}$. Each $b_i\ge a_i+2$, so $\Delta\ge2k$. Admissibility forces $k\le\max U$ and $k\le m-2-\max U$, hence $k\le p_{\max}$. The largest value of $\Delta$ with $|U|=k$ uses the $k$ smallest and $k$ largest indices, giving $f_m(k)$, which increases for $k\le p_{\max}$.
\end{proof}

Bond-dependent coupling therefore never enlarges the range of the invariant; it fills in values between the uniform ones. Table~\ref{tab:invariants} lists the sets of invariants for $5\le m\le8$ in the principal orientation. For $m\le4$, $\cV_{\mathrm{dir}}=\cV_{\mathrm{eq}}$, so $m=5$ is the smallest layer count with new values: $\cD_5=\{0,2,3\}$ against $\{0,3\}$ at equal coupling, giving $\mathfrak C=7$, $9$, and $10$. We checked Theorem~\ref{thm:configurations} numerically for $4\le m\le8$: $2\times10^4$ random monotone profiles and positive couplings per $m$ never produced a configuration outside the admissible set, and the construction realized every admissible pair.

\begin{table*}
\caption{Sets of invariants for descending North and ascending South potentials. $\cV_{\mathrm{eq}}$: equal coupling [Eq.~\eqref{eq:equal-invariants}]. $\cV_{\mathrm{sym}}$: reflection-symmetric bulks [Theorem~\ref{thm:symmetric}], which share a global gap by construction. $\cV_{\mathrm{dir}}$: all monotone pairs with direct gaps [Theorem~\ref{thm:configurations}]. $\cV_{\mathrm{glob}}$: values with a common global gap established here. A dash denotes a range of consecutive integers.}
\label{tab:invariants}
\begin{ruledtabular}
\setlength{\tabcolsep}{2pt}
\begin{tabular}{clllll}
$m$ & $\cV_{\mathrm{eq}}$ & $\cV_{\mathrm{sym}}$ & $\cV_{\mathrm{dir}}$ & $\cV_{\mathrm{glob}}$ & $\cV_{\mathrm{glob}}\setminus\cV_{\mathrm{eq}}$ \\
\hline
5 & 5, 8, 11 & 5, 8, 11 & 5, 7--11 & $=\cV_{\mathrm{dir}}$ & 7, 9, 10\\
6 & 6, 10, 12, 14, 16, 18 & 6, 8, \ldots, 18 (even) & 6, 8--16, 18 & $=\cV_{\mathrm{dir}}$ & 8, 9, 11, 13, 15\\
7 & 7, 12, 15, 17, 20, 23 & 7, 10, 12, 13, 15, 17, 18, 20, 23 & 7, 9--23 & 7, 9--20, 23 & 9--11, 13, 14, 16, 18, 19\\
8 & 8, 14, 18, 20, 24, 26, 28, 30, 32 & 8, 10, \ldots, 32 (even) & 8, 10--30, 32 & $\supseteq\cV_{\mathrm{sym}}$ & 10, 12, 16, 22\\
\end{tabular}
\end{ruledtabular}
\end{table*}

\subsection{Common global gaps}

The set $\cV_{\mathrm{dir}}$ describes band projectors. Channel counts are the elements of $\cV_{\mathrm{glob}}$, which require a common global gap [Eq.~\eqref{eq:intro-flow}]. With unrestricted offsets, this reduces to a property of each bulk alone.

\begin{proposition}[Gap alignment]
\label{prop:alignment}
A monotone pair admits offsets that give it a common global gap if and only if each bulk has a global gap. The offsets change neither the orientations nor the occupied configurations.
\end{proposition}

\begin{proof}
Adding a constant $c_B$ to all potentials of bulk $B$ shifts its spectrum rigidly by $c_B$. Choosing $c_N$ and $c_S$ so that both global gaps contain zero produces Eq.~\eqref{eq:global-gap}. The converse is immediate.
\end{proof}

Layer reversal preserves the radial spectrum, since it permutes the components of the real tridiagonal radial symbol. Hence the set $\cD^{\mathrm{glob}}_m\subseteq\cD_m$ of values of $\Delta$ realized by bulks with a global gap does not depend on orientation, and $\cV_{\mathrm{glob}}(m;\mathcal O)$ is Eq.~\eqref{eq:equal-invariants} with $f_m(p_B)$ ranging over $\cD^{\mathrm{glob}}_m$.

\begin{theorem}[Reflection-symmetric realizations]
\label{thm:symmetric}
For every $U\subseteq\{1,\ldots,p_{\max}\}$ the pair $(U,m-U)$ is admissible and is realized by a bulk with antisymmetric potentials, $u_{m+1-j}=-u_j$, and palindromic couplings, $\gamma_{m-j}=\gamma_j$. Such a bulk has a global gap around zero. Consequently
\begin{equation}
\Lambda(U)=\sum_{j\in U}(m-2j)\in\cD^{\mathrm{glob}}_m.
\label{eq:g}
\end{equation}
Conversely, every reflection-symmetric bulk with a direct gap has a configuration of this form.
\end{theorem}

\begin{proof}
Admissibility reads $2\max U+2\le m$, which is $\max U\le p_{\max}$. The construction in the proof of Theorem~\ref{thm:configurations}, with $C=m-U$ and an antisymmetric interpolation, is antisymmetric and palindromic. Lemma~\ref{lem:J} makes the spectrum symmetric about zero at every momentum, so the direct gap contains an interval around zero. In a symmetric bulk, $E^+_j<0$ if and only if $E^-_{m-j}>0$, which gives $C=m-U$.
\end{proof}

By the converse in Theorem~\ref{thm:symmetric}, $\cV_{\mathrm{sym}}(m;\mathcal O)$ is Eq.~\eqref{eq:equal-invariants} with $f_m(p_B)$ ranging over the values $\Lambda(U)$. Prefix sets $U=\{1,\ldots,p\}$ give $\Lambda=f_m(p)$, which proves the first two inclusions in Eq.~\eqref{eq:nesting}. The set $\cV_{\mathrm{sym}}$ adds values to $\cV_{\mathrm{eq}}$ for $m\ge6$; the smallest is $\mathfrak C=8$ at $m=6$, from $U_S=\{2\}$ and $U_N=\varnothing$. Proposition~\ref{prop:alignment} turns any bulk with a global gap into a building block for $\cV_{\mathrm{glob}}$. A lower bound on the global gap of one bulk for each value of $\Delta$ missing from the symmetric family (Appendix~\ref{app:cert}, Table~\ref{tab:asym}) establishes $\cD^{\mathrm{glob}}_m=\cD_m$ for $m=5$ and $6$. At $m=7$ no such bound is available yet for $\Delta=6$ and $7$, which leaves $\mathfrak C=21$ and $22$ open.

\subsection{Finding gapped examples}

The sets $\cD_m$ and $\cV_{\mathrm{dir}}$ follow from Theorem~\ref{thm:configurations} by enumeration. Which configurations also admit a global gap, and how large it can be, we found numerically in two ways.

\begin{figure*}
\includegraphics[width=\textwidth]{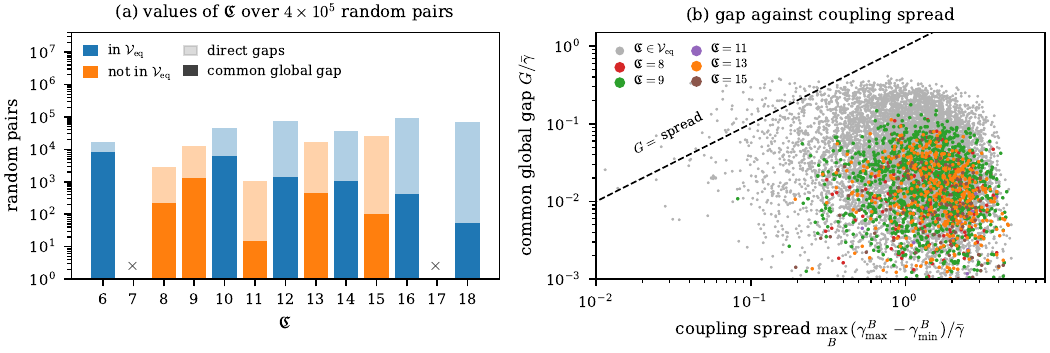}
\caption{Random search at $m=6$. (a) Values of $\mathfrak C=6+\Delta_N+\Delta_S$ over $4\times10^5$ random pairs of sampled bulks in the field-reversal orientation, on a logarithmic scale. Light bars count pairs with direct gaps, dark bars pairs with a common global gap; blue marks values in $\cV_{\mathrm{eq}}$ and orange values outside it. Crosses mark values that never occur. (b) Common global gap against coupling spread, both divided by the larger of the two mean couplings, for pairs with a common global gap. Gray: values in $\cV_{\mathrm{eq}}$ (random subsample); colors: values outside $\cV_{\mathrm{eq}}$. Proposition~\ref{prop:spread} requires the colored points to lie below the dashed line $G=$ spread. Global gaps are estimated on a radial grid.}
\label{fig:search}
\end{figure*}

\emph{Random sampling.} We draw ascending bulks with independent random potential increments and couplings,
\begin{lstlisting}
def sample_bulk(m, rng):
    du = np.exp(rng.normal(0, 1, m-1) + rng.normal(0, 0.7))
    u = np.cumsum(np.r_[0, du]); u -= u.mean()
    s = rng.uniform(0, 1.2)     # spread of the log-couplings
    g = np.exp(rng.normal(0, 0.8) + s*rng.normal(0, 1, m-1))
    return u, g
\end{lstlisting}
For each sample we compute $\Delta$ from the occupied weight at zero momentum (Theorem~\ref{thm:weights}) and estimate the global gap on a radial grid of 220 points. All of $8\times10^4$ six-layer samples have a direct gap, and $\Delta$ takes exactly the values $\{0,2,3,4,6\}$ of $\cD_6$. Each value also occurs with a global gap, including the two outside the equal-coupling set, $\Delta=2$ (152 samples) and $\Delta=3$ (900 samples); this checks $\cD^{\mathrm{glob}}_6=\cD_6$ independently of Table~\ref{tab:asym}. By Proposition~\ref{prop:alignment}, a random pair's common global gap, with offsets chosen freely, is the smaller of its two global gaps. Pairing samples at random gives the distribution of $\mathfrak C$ in Fig.~\ref{fig:search}(a). It is exactly $\cV_{\mathrm{dir}}(6)=\{6,8\text{--}16,18\}$; the values 7 and 17 never occur, and values outside $\cV_{\mathrm{eq}}$ are rarer, especially with a common global gap.

\emph{Targeted search.} For a prescribed pair $(U,C)$ we maximize the global gap over the logarithms of the potential increments and couplings with the Nelder--Mead method, starting from the construction in the proof of Theorem~\ref{thm:configurations}. We then bound the global gap of the best bulks from below (Appendix~\ref{app:cert}). Table~\ref{tab:asym} lists them, and wall ii below uses the five-layer one.

\subsection{Examples}

\begin{table}
\caption{Bulk pairs used for the interface calculations, in units of $E_0$ ($\hbar v=1$). Wall i: $u^S_j=-u^N_j=(j-2)E_0$, $\gamma_j=E_0$. Walls ii--iv: see text. Walls v and vi: Eq.~\eqref{eq:family} at $g=1.8$ with gate slope $s=0.5$ and $s=1$. Wall vii: see text (sign reversal, South descending). Window: interval avoided by both bulk spectra, from the lower-bound method of Appendix~\ref{app:cert}. $\bar\gamma$: mean coupling of the bulk with larger mean. $r_\gamma$: coupling contrast, Eq.~\eqref{eq:contrast}. SF: signed crossing count of the interface spectrum (Sec.~\ref{sec:numerics}), identical for sharp and smooth walls.}
\label{tab:examples}
\begin{ruledtabular}
\begin{tabular}{lccccccc}
Wall & $m$ & $\mathfrak C$ & Window & $G/\bar\gamma$ & $r_\gamma$ & SF\\
\hline
i & 3 & 3 & $\pm0.320$ & 0.640 & 0 & 3\\
ii & 5 & 9 & $(-0.442,0.095)$ & 0.286 & 0.33 & 9\\
iii & 5 & 7 & $(-0.370,0.095)$ & 0.233 & 0.33 & 7\\
iv & 6 & 8 & $\pm0.0184$ & 0.020 & 0.60 & 8\\
v & 8 & 8 & $\pm0.409$ & 0.273 & 0.40 & 8\\
vi & 8 & 16 & $\pm0.105$ & 0.070 & 0.40 & 16\\
vii & 6 & $-10$ & $\pm0.0557$ & 0.177 & 0.59 & $-10$\\
\end{tabular}
\end{ruledtabular}
\end{table}

Table~\ref{tab:examples} collects the bulk pairs whose interface spectra we compute. All gap windows come from the lower-bound argument of Appendix~\ref{app:cert}.

\begin{figure}
\includegraphics[width=\columnwidth]{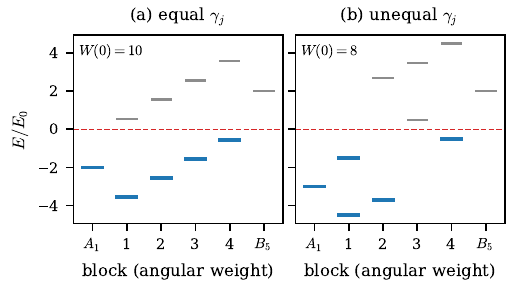}
\caption{Zero-momentum levels of five-layer bulks, grouped by block ($A_1$, coupled pairs $1$--$4$, $B_5$). A block's angular weight equals its label. Thick bars are occupied; the dashed line marks the gap center. (a) $\bm u=(-2,-1,0,1,2)E_0$, $\gamma_j=2E_0$: the occupied levels form the equal-coupling prefix, $W(0)=10$. (b) South bulk of wall ii: both levels of pair 1 are occupied and both levels of pair 3 are unoccupied, $W(0)=8$, which equal coupling forbids.}
\label{fig:levels}
\end{figure}

\begin{figure}
\includegraphics[width=\columnwidth]{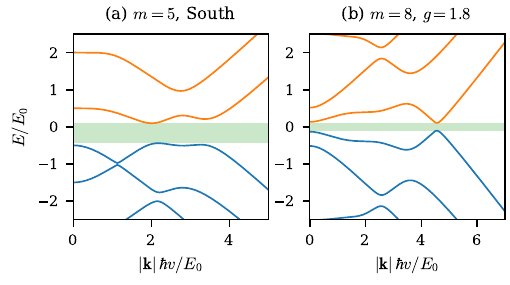}
\caption{Radial bulk spectra (lower bands blue, upper bands orange) with the common global gap shaded (lower bound, Appendix~\ref{app:cert}). (a) South bulk of wall ii; the North bulk has the same radial spectrum. (b) Eight-layer bulk of wall vi ($g=1.8$, $s=1$). The gap is limited by a finite-momentum minimum near $|\bm k|\simeq4.55E_0/\hbar v$.}
\label{fig:bulk}
\end{figure}

\emph{Five layers.} The South bulk $\bm u^S=(-3,-3,2,2,2)E_0$, $\bm\gamma^S=(1.5,2,1.5,2.5)E_0$ has both levels of pair 1 occupied and both levels of pair 3 unoccupied: $U=\{1\}$, $C=\{3\}$, $W_S=8$ [Fig.~\ref{fig:levels}(b)]. Equal coupling forbids this configuration. Its radial spectrum has a global gap [Fig.~\ref{fig:bulk}(a)]. Pairing it with its layer reversal gives $W_N=17$ and $\mathfrak C=9$ (wall ii). Pairing it with the linear bulk $\bm u^N=(2,1,0,-1,-2)E_0$, $\bm\gamma^N=2E_0$ gives $\mathfrak C=7$ (wall iii). Neither value is attainable at equal coupling for any orientation pair, since the equal-coupling curvatures are $I_B\in\{\pm2.5,\pm5.5\}$. Wall ii changes the coupling pattern across the interface, so it is not purely electrostatic. Its potential profile also concentrates a drop of $5E_0$ in one gallery (Sec.~\ref{sec:discussion}).

\emph{Six layers, gate-defined.} With the same couplings $\bm\gamma=(3,1,1.4,1,3)E_0$ on both sides and linear potentials $u^S_j=1.5E_0(j-\tfrac72)$ and $u^N_j=-0.5E_0(j-\tfrac72)$, the South bulk has $U=\{2\}$ and the North bulk is in the reference state, giving $\mathfrak C=8$ (wall iv). This wall is defined entirely by gates, but its gap is small.

\emph{Six layers, sign reversal.} A wall whose North bulk is the layer reversal of its South bulk has $\Delta_N=\Delta_S=\Delta$ and $|\mathfrak C|=m+2\Delta$. At equal coupling this allows only $|\mathfrak C|\in\{6,14,18\}$ for $m=6$. Wall vii uses the antisymmetric potentials $\bm u^N=-\bm u^S=5\,(-0.1117,-0.0918,-0.0691,0.0691,0.0918,0.1117)E_0$ and the palindromic couplings $\bm\gamma=(0.7374,0.3336,1,0.3336,0.7374)E_0$ on both sides. Its bulks have $U=\{2\}$ and $C=\{4\}$, so $\Delta=2$, and South is descending, so $\mathfrak C=-10$. This is the six-layer counterpart of wall vi below.

\begin{figure}
\includegraphics[width=\columnwidth]{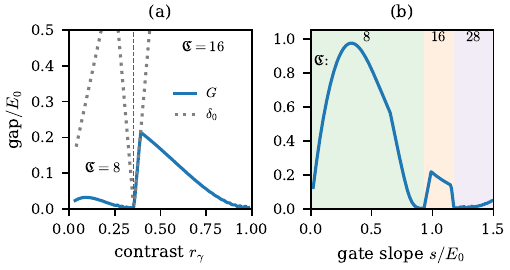}
\caption{(a) Along the eight-layer family at $s=1$: common global gap $G$ and zero-momentum gap $\delta_0$ against the contrast $r_\gamma=1-g/3$. The dashed line marks the transition at $r_\gamma=1-\sqrt{15}/6$. (b) At the fixed coupling pattern $\bm\gamma(1.8)$: common gap against the gate slope $s$, with the windows $\mathfrak C=8$, $16$, and $28$ shaded.}
\label{fig:tradeoff}
\end{figure}

\emph{Eight layers, gate-defined.} Take linear potentials $u^S_j=sE_0(j-\tfrac92)=-u^N_j$, in which neighboring layers differ in potential by the same step $sE_0$; we call $s$ the gate slope. Take also the palindromic couplings
\begin{equation}
\bm\gamma(g)=E_0(g_{\mathrm s},g,g_{\mathrm s},g_{\mathrm s},g_{\mathrm s},
g,g_{\mathrm s}),\qquad g_{\mathrm s}=\frac{21-2g}{5},
\label{eq:family}
\end{equation}
whose mean is fixed at $\gref=3E_0$. At $s=1$ the South bulk has $U=\{2\}$ and $C=\{6\}$ for $0<g<\sqrt{15}/2$, the North bulk mirrors it, and $\mathfrak C=8+2\Lambda(\{2\})=16$. This value is absent from $\cV_{\mathrm{eq}}$ at eight layers for every orientation pair, since $I_B\in\{\pm4,\pm10,\pm14,\pm16\}$ there. At $g=\sqrt{15}/2$ the gap closes at zero momentum and $\mathfrak C$ drops to 8. For $g$ close to zero the gap is limited instead by a finite-momentum minimum [Fig.~\ref{fig:bulk}(b)]. The best compromise along the family is $G\simeq0.213E_0$ at $g\simeq1.83$ [Fig.~\ref{fig:tradeoff}(a) and Table~\ref{tab:family}].

At the fixed coupling pattern $\bm\gamma(1.8)$, the gate slope $s$ alone selects the invariant [Fig.~\ref{fig:tradeoff}(b)]: $\mathfrak C=8$ for $s<2g/\sqrt{15}\simeq0.930$, $\mathfrak C=16$ for $0.930<s<2g_{\mathrm s}/\sqrt{35}\simeq1.176$, and $\mathfrak C=28$ beyond, where $(U,C)=(\{1,2\},\{6,7\})$. Walls v and vi are the same structure at two gate strengths. They carry 8 and 16 co-propagating channels per valley, with common global gaps of at least $0.819E_0$ and $0.210E_0$.

\subsection{Necessary coupling spread}
\label{sec:contrast}

Call a bulk \emph{uniform} if all its couplings are equal. By Lemma~\ref{lem:prefix}, a uniform bulk has $\Delta=f_m(p)$ for some $p\le p_{\max}$, whatever the value of its coupling. A pair of uniform bulks therefore has $\mathfrak C\in\cV_{\mathrm{eq}}$ even when the two bulks have different coupling values. A new invariant thus requires that at least one bulk be far from uniform, and the gap sets how far.

\begin{proposition}[Necessary spread]
\label{prop:spread}
Let $\kappa_B=\gamma^B_{\max}-\gamma^B_{\min}$ and $\kappa=\max_B\kappa_B$. If a monotone pair has a common global gap $G$ and $\mathfrak C\notin\cV_{\mathrm{eq}}$, then
\begin{equation}
\kappa\ge G.
\label{eq:spread}
\end{equation}
\end{proposition}

\begin{proof}
Let $\mu_B=(\gamma^B_{\max}+\gamma^B_{\min})/2$ and, for $t\in[0,1]$, let $H_B(\bm k;t)$ be the bulk Hamiltonian with couplings $\gamma^B_j(t)=(1-t)\gamma^B_j+t\mu_B$. The difference does not depend on $\bm k$ and acts only on the coupled pairs,
\begin{equation}
H_B(\bm k;t)-H_B(\bm k;0)
=\bigoplus_{j=1}^{m-1}t\bigl(\mu_B-\gamma^B_j\bigr)\sigma_1 ,
\end{equation}
where block $j$ acts on $\operatorname{span}\{B_j,A_{j+1}\}$; it vanishes on $A_1$ and $B_m$. Hence, for every $\bm k$,
\begin{equation}
\bigl\|H_B(\bm k;t)-H_B(\bm k;0)\bigr\|=t\max_j\bigl|\gamma^B_j-\mu_B\bigr|
=\tfrac t2\kappa_B .
\end{equation}
By Weyl's inequality~\cite{Bhatia}, $|e^B_\ell(r;t)-e^B_\ell(r;0)|\le\tfrac t2\kappa_B$ for all $\ell$ and $r$, so with $E_{\mathrm v}(t)$, $E_{\mathrm c}(t)$, $G(t)$ as in Eq.~\eqref{eq:global-gap},
\begin{equation}
E_{\mathrm v}(t)\le E_{\mathrm v}+\tfrac t2\kappa,\qquad
E_{\mathrm c}(t)\ge E_{\mathrm c}-\tfrac t2\kappa,\qquad
G(t)\ge G-t\kappa .
\end{equation}
Suppose $\kappa<G$. Then $G(t)>0$ on $[0,1]$, so both bulks keep a direct gap, and $W_B(t)=\Tr\bigl(DP_B(\bm 0;t)\bigr)$ is continuous in $t$ and integer-valued (Theorem~\ref{thm:weights}), hence constant. Therefore $\mathfrak C(t)=W_N(t)-W_S(t)$ is constant. At $t=1$ both bulks are uniform, so $\mathfrak C(1)\in\cV_{\mathrm{eq}}$ by Lemma~\ref{lem:prefix}, a contradiction.
\end{proof}

In relative terms, let $\bar\gamma_B$ be the mean coupling of bulk $B$ and
\begin{equation}
r_\gamma=\max_{B,j}\Bigl|\frac{\gamma^B_j}{\bar\gamma_B}-1\Bigr|,
\label{eq:contrast}
\end{equation}
the largest fractional deviation of a bond from its bulk's mean. Since $\kappa_B\le2r_\gamma\bar\gamma_B$, Eq.~\eqref{eq:spread} gives $r_\gamma\ge G/(2\max_B\bar\gamma_B)$.

The bound is necessary but far from sharp. In the random search at $m=6$ [Fig.~\ref{fig:search}(b)], all 2097 pairs with a common global gap and $\mathfrak C\notin\cV_{\mathrm{eq}}$ have $G\le0.27\kappa$, while pairs with $\mathfrak C\in\cV_{\mathrm{eq}}$ reach gaps several times their spread. For wall vi the bound asks for a spread of at least $0.21E_0$; the construction uses $1.68E_0$. Table~\ref{tab:family} shows how the gap of the eight-layer family depends on its contrast.

\begin{table}
\caption{The eight-layer family, Eq.~\eqref{eq:family}, at $s=1$ and fixed mean coupling $3E_0$. $\delta_0=e_{m+1}(0)-e_m(0)$ is the zero-momentum gap. $G$ is a lower bound on the common global gap (grid spacing $\delta r=5\times10^{-4}E_0$), in units of $E_0$. All rows have $\mathfrak C=16$.}
\label{tab:family}
\begin{ruledtabular}
\begin{tabular}{ccccc}
$g$ & $r_\gamma$ & $\delta_0/E_0$ & $G/E_0$ & $G/\gref$\\
\hline
0.5 & 0.833 & 2.062 & 0.026 & 0.009\\
1.0 & 0.667 & 1.666 & 0.090 & 0.030\\
1.5 & 0.500 & 0.838 & 0.167 & 0.056\\
1.8 & 0.400 & 0.264 & 0.210 & 0.070\\
1.826 & 0.391 & 0.214 & 0.213 & 0.071\\
1.9 & 0.367 & 0.071 & 0.070 & 0.023\\
\end{tabular}
\end{ruledtabular}
\end{table}

\section{Interface spectra}
\label{sec:numerics}

\subsection{Method}

Equation~\eqref{eq:intro-flow} predicts the signed number of interface channels from bulk data alone. We test it by computing that number directly from the interface operator, for each wall of Table~\ref{tab:examples}. The measured quantity is $\SF_{k_y\uparrow}(H_I(k_y)-E_F)$: the eigenvalues of the one-dimensional operator $H_I(k_y)$ of Eq.~\eqref{eq:interface} that lie in the common global gap are located as functions of $k_y$, and their crossings of a fixed $E_F$ are counted with the sign of the slope. The bulk side, $\mathfrak C=W_N-W_S$, uses only the zero-momentum levels (Theorem~\ref{thm:weights}); the two computations share no code.

Potentials and couplings interpolate between the bulks through $(1-\mathcal S)\bm u^S+\mathcal S\bm u^N$ and $(1-\mathcal S)\bm\gamma^S+\mathcal S\bm\gamma^N$, where $\mathcal S(x)$ is a $C^\infty$ step that equals $0$ for $x\le-w$ and $1$ for $x\ge w$, with $w=2\hbar v/E_0$. We divide $-w\le x\le w$ into $2n$ slices of equal width $\Delta x=w/n$, with $n=20$ and $40$, and hold $\mathcal S$ constant within each slice, equal to its value at the slice's midpoint. We also use a sharp wall ($n=0$). Outside $|x|\le w$ the coefficients are exactly constant, so the asymptotic data below are exact.

\subsection{Algorithm}

An eigenvalue $E$ of $H_I(k_y)$ is an energy at which a solution decays on both sides. Writing $H_I(k_y)\psi=E\psi$ as $\partial_x\psi=A(x)\psi$ with
\begin{equation}
A=-i\Sigma_1\bigl(k_y\Sigma_2+V(x)-E\bigr),
\label{eq:transfer}
\end{equation}
the computation has four steps.
\begin{enumerate}\setlength{\itemsep}{1pt}\setlength{\parskip}{0pt}
\item \emph{Boundary data.} For $E$ in the common gap, each bulk matrix $A_B$ has $m$ eigenvalues with positive and $m$ with negative real part. An ordered Schur form gives orthonormal bases $Q_L$ of the South subspace that decays toward $x\to-\infty$ and $Q_R$ of the North subspace that decays toward $x\to+\infty$.
\item \emph{Propagation.} Within a slice $A$ is constant, so a frame $X$ is advanced exactly: $X(x+\Delta x)=e^{A\Delta x}X(x)$. $Q_L$ is carried this way across the South half of the wall, and $Q_R$ across the North half with $e^{-A\Delta x}$. After every slice the frame is replaced by the orthonormal factor of its QR factorization~\cite{Humpherys,Ledoux}. This keeps its span and prevents its columns from collapsing onto the fastest-growing mode.
\item \emph{Matching.} At $x=0$, $E$ is an eigenvalue exactly when the two $m$-dimensional subspaces $\mathcal L=\operatorname{span}Q_L$ and $\mathcal R=\operatorname{span}Q_R$ intersect. Let $[\,Q_L\;Q_R\,]$ be the $2m\times2m$ matrix whose columns are those of $Q_L$ followed by those of $Q_R$. Its rank is $\dim(\mathcal L+\mathcal R)$, so $\dim(\mathcal L\cap\mathcal R)=2m-\operatorname{rank}[\,Q_L\;Q_R\,]$: the subspaces intersect exactly when the smallest singular value $\sigma_{\min}$ of $[\,Q_L\;Q_R\,]$ vanishes, and the number of vanishing singular values is the multiplicity of the eigenvalue. With $c$ the largest cosine of the principal angles between the subspaces, $\sigma_{\min}=\sqrt{1-c}$.
\item \emph{Counting.} At a fixed $E_F$ in the window of Table~\ref{tab:examples}, offset from its center by $1.37\%$ of its width to avoid coincidences, the zeros of $k_y\mapsto\sigma_{\min}(k_y,E_F)$ are located and refined. The sign of $dE/dk_y$ at each zero is read off from the eigenvalue refined at $k_y\pm2\times10^{-5}$, and the signs are summed.
\end{enumerate}
Every point $(k_y,E)$ requires the same fixed number of matrix exponentials and QR factorizations, and the points are independent. The computation is therefore compiled once and evaluated over a whole grid in parallel. In JAX, steps 1--3 and the grid evaluation read:
\begin{lstlisting}
def sigma_min(ky, E):                # one (ky, E) point
    A = lambda V: -1j*S1 @ (ky*S2 + V - E*I)
    QL = decaying(A(VS), left=True)  # 2m x m frames
    QR = decaying(A(VN), left=False)
    fwd = lambda Q, V: (qr(expm( A(V)*dx) @ Q)[0], None)
    bwd = lambda Q, V: (qr(expm(-A(V)*dx) @ Q)[0], None)
    QL, _ = jax.lax.scan(fwd, QL, V_south)  # -w to 0
    QR, _ = jax.lax.scan(bwd, QR, V_north)  # +w to 0
    return svd(jnp.hstack([QL, QR]), compute_uv=False)[-1]

# all (ky, E) points in parallel, compiled once
grid = jax.jit(jax.vmap(jax.vmap(sigma_min, (None, 0)),
                        (0, None)))
scores = grid(ks, Es)
\end{lstlisting}
Here \texttt{decaying} returns an orthonormal basis of the eigenvectors of its argument whose eigenvalues have positive (\texttt{left}) or negative real part, and \texttt{V\_south}, \texttt{V\_north} hold the matrices $V$ of the slices on either side of $x=0$. On one CPU core an eight-layer wall takes $0.37$~ms per point for a sharp wall and $5.5$~ms per point with 40 slices.

\subsection{Error control}

\emph{Transfer.} For a piecewise-constant profile the slice propagators are exact, so the only error in steps 1--3 is floating-point rounding. Across one slice the frame can be stretched by at most $e^{\Lambda\Delta x}$, where $\Lambda$ is the spread of the real parts of the eigenvalues of $A$. Over the whole scan window and for all walls here $\Lambda\le18E_0/\hbar v$, growing roughly as $2|k_y|$, so with $\Delta x\le0.1\hbar v/E_0$ the stretch per slice is at most $e^{1.8}\approx6$. Re-orthonormalizing after every slice keeps the frames well conditioned, and all arithmetic is in double precision.

\emph{Profile.} Sampling a smooth $\mathcal S$ at slice midpoints changes the wall profile, not the bulks, so it moves the crossings but cannot change their signed number. The crossing positions converge at second order in $\Delta x$: for wall i, successive differences as $n$ doubles from 5 to 80 are $1.2\times10^{-3}$, $3.1\times10^{-4}$, $7.9\times10^{-5}$, and $1.9\times10^{-5}$ (in units of $E_0/\hbar v$), so at $n=20$ the positions are accurate to about $10^{-4}$. The count is 3 for every $n$.

\emph{Search window.} No crossing exists outside $|k_y|\le K$ with $K=\max_B\|V_B\|+|E_F|$. Indeed $(-i\Sigma_1\partial_x+k_y\Sigma_2)^2=-\partial_x^2+k_y^2$, so an eigenvalue $E$ at $k_y$ requires $|k_y|\le|E|+\sup_x\|V(x)\|$, and the interpolation gives $\sup_x\|V(x)\|\le\max_B\|V_B\|$. We scan $[-K-2,K+2]$.

\emph{Resolution.} The scan uses $N$ points ($N=3000$ and $6000$ for sharp walls, $2500$--$3000$ for smooth walls). Every cell whose minimum of $\sigma_{\min}$ is below $0.3$ is rescanned on 400 points, which resolves crossings about $10^{-5}E_0/\hbar v$ apart; the closest pair found, in wall ii, is $1.6\times10^{-3}$ apart. A zero is accepted when the refined minimum is below $10^{-4}$ and below $10^{-3}$ times its values at $k_y\pm10^{-4}$. The relative test is needed because at large $|k_y|$ the decay rates of the bulk modes differ by up to a factor of 50, which limits $\sigma_{\min}$ at a genuine crossing to about $10^{-6}$; the neighbors of such a zero are near $10^{-2}$, while near-misses have no such contrast. Since $\partial H_I/\partial k_y=\Sigma_2$ has norm one, $|dE/dk_y|\le1$, so the energy refinement in step 4 needs a window of only $\pm3\times2\times10^{-5}$. The gap windows themselves are the lower bounds of Appendix~\ref{app:cert}. Doubling $N$, and changing $n$ between 0, 20, and 40, leaves every count unchanged.

\subsection{Results}

\begin{figure}
\includegraphics[width=\columnwidth]{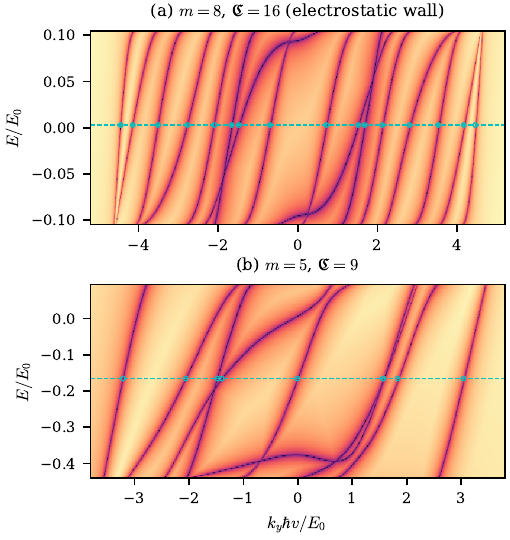}
\caption{Interface spectra of the sharp walls vi and ii. Horizontal axis: momentum $k_y$ along the wall; vertical axis: energy $E$, restricted to the common global gap of Table~\ref{tab:examples}; color: $\log_{10}\sigma_{\min}$, which tends to $-\infty$ on an interface eigenvalue branch (dark curves). The dashed line is $E_F$, and circles mark the refined crossings entering the count. (a) Wall vi: 16 crossings, all with positive slope. (b) Wall ii: 9 crossings, all with positive slope; two lie only $1.6\times10^{-3}$ apart near $k_y\simeq1.57$.}
\label{fig:interface}
\end{figure}

Table~\ref{tab:examples} lists the results, and Fig.~\ref{fig:interface} shows two of the spectra. For every wall the signed count equals $\mathfrak C$, for sharp walls at two scan resolutions and for smooth walls. The uniform three-layer wall fixes the sign convention: its descending North and ascending South give $\SF=+3=I_N-I_S$. Every crossing has multiplicity one, and all crossings of a given wall have the sign of $\mathfrak C$, so in these examples $|\mathfrak C|$ is also the number of channels moving in one direction. Two independent implementations, one in SciPy and one in JAX, give identical counts.

\section{Discussion}
\label{sec:discussion}

The occupied-weight formula separates the two controls cleanly. At equal coupling, the ordering of the upper and of the lower levels survives any monotone potential, so screening moves transitions but cannot create invariant values. Bond selectivity breaks that ordering. It allows a configuration in which a weakly coupled pair keeps both levels on one side of the gap while its strongly coupled neighbors straddle it. The admissibility condition $\max U+2\le\min C$ is the only remaining constraint, so $\cV_{\mathrm{dir}}$ is determined combinatorially.

\emph{Connection to experiment.} In a split-gate device like those of Refs.~\cite{LiNN,Huang2024}, a wall with invariant $\mathfrak C$ carries $|\mathfrak C|$ channels per valley and spin in each direction. Without intervalley scattering its two-terminal conductance is $2|\mathfrak C|e^2/h$, from two spins and one valley per direction. This gives the measured $4e^2/h$ for the bilayer, and $2m\,e^2/h$ for an $m$-layer field reversal at equal coupling, the value found in tight-binding transport calculations for rhombohedral tetralayers and pentalayers~\cite{ShenHou}. Those calculations also find the kink states concentrated on the outermost layers, which makes them sensitive to disorder there~\cite{ShenHou}. No domain-wall experiment beyond the bilayer has been reported. The walls predicted here would add conductance values that an equal-coupling stack cannot show at a field reversal; for instance, the eight-layer stack of walls v and vi switches from $16$ to $32\,e^2/h$ as the gate strength increases.

Physically, this requires the following. A new configuration needs a pair whose level splitting, set by $\gamma_j$, is smaller than the offset of its center from the gap, which is set by the potential drop. Selected galleries must therefore be weakly coupled compared with the local potential difference, while their neighbors remain strongly coupled. To attach numbers, identify the mean coupling of wall vi, $3E_0$, with $\gamma_1\simeq0.38$~eV~\cite{CastroNeto,McCannKoshino}, which gives $E_0\simeq0.13$~eV. The wall then needs a potential step of about 0.13~eV per layer and has $G\simeq27$~meV. For comparison, the gap that a perpendicular field opens in bilayer graphene can be tuned up to about 250~meV~\cite{ZhangGap}, the scale of the bulk gaps behind the measured kink states. No gap separates the two valleys; they are kept apart by their separation in momentum, and $G$ is the energy window in which $E_F$ must lie. A gap of 27~meV corresponds to about 300~K, so at the temperatures of the bilayer experiments~\cite{Huang2024} it would not be the limiting scale. The gap minimum sits at $|\bm k|\simeq0.9$~nm$^{-1}$ ($\hbar v\simeq0.66$~eV\,nm), where the omitted trigonal-warping term $\hbar v_3k$, with $v_3\approx0.1v$~\cite{McCannKoshino}, is of order 60~meV. The five-layer walls have larger relative gaps, $G/\bar\gamma\simeq0.29$, but wall ii concentrates a potential drop of about 1~eV in a single gallery, far beyond what screened devices produce. The examples therefore demonstrate the mechanism; they are not device designs. Converting a coupling pattern into intercalant staging, gallery spacing, or a heterostructure requires a microscopic model. A realistic design also needs the common gap rechecked with trigonal warping and remote interlayer couplings, which break the continuous rotational covariance used here but preserve the invariant while the common gap stays open.

\emph{Built-in level offsets.} The new configurations need large, layer-selective level offsets, which our examples obtain from strong gate fields. A layered material can supply such offsets without gates. Monolayer honeycomb SiC has recently been grown epitaxially~\cite{Polley}. Its two sublattices are inequivalent, which in the continuum description adds a sublattice-dependent energy to that layer; like a perpendicular field, such a term gaps both valleys with opposite Berry curvature. The occupied-weight formula extends unchanged to sublattice-dependent energies: a diagonal term commutes with $e^{i\theta D}$ and leaves the principal symbol untouched, so Theorem~\ref{thm:weights} holds with the energies of $A_1$ and $B_m$ as the unpaired levels and the energies of $B_j$ and $A_{j+1}$ on the diagonal of pair $j$. The classification of Secs.~\ref{sec:equal} and~\ref{sec:unequal}, which assumes one potential per layer, would have to be redone. Stacks that interleave graphene layers, which carry the winding, with SiC-like layers, which supply the offsets, are therefore a natural place to look for the configurations of Sec.~\ref{sec:unequal}. Two caveats apply. The lattice constant of honeycomb SiC, about 3.1~\AA~\cite{Polley}, exceeds graphene's 2.46~\AA\ by about 27\%, so such a stack is not a commensurate ABC crystal and needs a moir\'e description; and its interlayer couplings and velocities would have to come from a microscopic calculation.

Our results have several limitations. Interactions, which drive spontaneous symmetry breaking in rhombohedral multilayer graphene~\cite{ZhangSQH,Han,Lu}, are absent. The invariant is valley-resolved, and its transport meaning assumes negligible intervalley scattering. Our wall profiles for the numerics are piecewise constant, which lies outside the smoothness assumed in the correspondence theorem; the agreement of all counts with $\mathfrak C$ is consistent evidence, not a proof for such profiles. On the classification side, $\cV_{\mathrm{glob}}$ is complete here only for $m\le6$. The values $\Delta=6$ and $7$ at $m=7$, and the odd values of $\mathfrak C$ at $m=8$, still require single-bulk gaps with a lower bound. A general criterion for when an admissible pair admits a global gap would settle $\cV_{\mathrm{glob}}$ for all $m$.

\section*{Data availability}

The code that computes occupied weights, gap lower bounds, and interface spectral flow, together with the data underlying the figures and tables, is available at \url{https://github.com/YuelongLi/RHG_Spectral_Indices}.

\begin{acknowledgments}
The author thanks Guillaume Bal and Matthew Frazier for guidance, helpful
discussions, and feedback on this work.
\end{acknowledgments}

\appendix

\section{Proof of Theorem~\ref{thm:weights}}
\label{app:weights}

We prove Theorem~\ref{thm:weights} using only finite-dimensional linear algebra and calculus. All matrices are $2m\times2m$, $\Tr$ is the matrix trace, and $\|\cdot\|$ is the operator norm. We use two standard facts:
\begin{itemize}\setlength{\itemsep}{1pt}\setlength{\parskip}{0pt}
\item \emph{Weyl's inequality}~\cite{Bhatia}: for Hermitian $A$ and $B$ with eigenvalues in increasing order, $|\lambda_k(A+B)-\lambda_k(A)|\le\|B\|$.
\item \emph{Resolvent formula}: if a closed contour $\Gamma$ in $\mathbb C$ encloses exactly the eigenvalues $\lambda_1,\ldots,\lambda_m$ of a Hermitian $A$, the projector onto their eigenspace is $P=\frac1{2\pi i}\oint_\Gamma(z-A)^{-1}dz$.
\end{itemize}

\subsection{Results used without proof}

We take Eq.~\eqref{eq:intro-flow} from Refs.~\cite{BalInterface,QuinnBal,FrazierBal} without reproving it. For orientation: with $\varphi$ the energy switch of Eq.~\eqref{eq:intro-sigma}, the operator $\mathcal U=e^{2\pi i\varphi(H_I)}$ acts as the identity on states whose energy lies outside the window where $\varphi$ rises. On an interface branch $E_n(k_y)$ that crosses the window, its eigenvalue $e^{2\pi i\varphi(E_n)}$ winds once around the unit circle as $k_y$ varies. Restricting $\mathcal U$ to the half plane $y>0$, i.e., $T=Q_y\mathcal UQ_y|_{\operatorname{Ran}Q_y}$ with $Q_y=\mathbf 1_{(0,\infty)}(y)$, turns this winding into a Fredholm index, and $\Ind T=2\pi\sigma_I=\mathfrak C$. For general elliptic interface operators with a common gap, this already shows that $\mathfrak C$ is an integer. Theorem~\ref{thm:weights} gives an independent proof for the present model that needs only a direct gap in each bulk.

\subsection{Direct gap}

\begin{lemma}[Zero-momentum gap criterion]
\label{lem:zero-gap}
For positive couplings, $e^B_m(0)<e^B_{m+1}(0)$ if and only if $\delta_B>0$.
\end{lemma}

\begin{proof}
At polar angle zero the symbol is $H_B(r,0)=r\Sigma_1+V_B$, a real symmetric tridiagonal matrix whose off-diagonal entries alternate between $r$ and $\gamma_j$. For $r>0$ all of them are nonzero. The eigenvalue equation then determines every component of an eigenvector from the first one, row by row, so each eigenspace is one-dimensional and the spectrum is simple; in particular $e^B_{m+1}(r)-e^B_m(r)>0$ for $r>0$. The matrix $r\Sigma_1$ has eigenvalue $-r$ and $+r$, each $m$ times, so Weyl's inequality gives $e^B_m(r)\le-r+\|V_B\|$ and $e^B_{m+1}(r)\ge r-\|V_B\|$, and the separation exceeds 1 for $r>\|V_B\|+\tfrac12$. On the remaining interval $[0,\|V_B\|+\tfrac12]$ the separation is continuous and positive if it is positive at $r=0$, so it has a positive minimum. The converse is immediate.
\end{proof}

\subsection{The projector is smooth and rotation covariant}

Fix a bulk and write $P(\bm k)$ for $P_B(\bm k)$. Near any $\bm k_0$, choose $\Gamma$ enclosing the $m$ lowest eigenvalues of $H_B(\bm k_0)$ and no others; the direct gap makes this possible. Eigenvalues depend continuously on $\bm k$, so the same $\Gamma$ works for all $\bm k$ near $\bm k_0$. The entries of $(z-H_B(\bm k))^{-1}$ are, by Cramer's rule, ratios of polynomials in $z$, $k_x$ and $k_y$ with a nonvanishing denominator on $\Gamma$, so the resolvent formula makes $P$ infinitely differentiable. Degeneracies among the $m$ lowest eigenvalues play no role: $P$ is defined without choosing eigenvectors.

In polar coordinates write $P(\theta,r):=P(r\cos\theta,r\sin\theta)$ and $U(\theta)=e^{i\theta D}$. Eq.~\eqref{eq:covariance} says $H_B(\theta,r)=U(\theta)H_B(0,r)U(\theta)^\dagger$. Conjugation by a unitary maps the resolvent of one matrix to the resolvent of the other, so the resolvent formula gives
\begin{equation}
P(\theta,r)=U(\theta)P(0,r)U(\theta)^\dagger,\qquad
\partial_\theta P=i[D,P],
\label{eq:P-covariance}
\end{equation}
where the second identity follows from $\partial_\theta U=iDU$. Since $D$ commutes with $U(\theta)$, the weight $W_B(r)=\Tr\bigl(DP(\theta,r)\bigr)=\Tr\bigl(DP(0,r)\bigr)$ does not depend on $\theta$.

\subsection{The curvature in polar coordinates}

Write $P_x=\partial_{k_x}P$, $P_y=\partial_{k_y}P$, $P_r=\partial_rP$ and $P_\theta=\partial_\theta P$. The chain rule gives $P_r=\cos\theta\,P_x+\sin\theta\,P_y$ and $P_\theta=r(-\sin\theta\,P_x+\cos\theta\,P_y)$, hence
\begin{equation}
[P_r,P_\theta]=r\bigl(\cos^2\theta+\sin^2\theta\bigr)[P_x,P_y]=r[P_x,P_y].
\end{equation}
With $d^2k=r\,dr\,d\theta$, the curvature integrand over an annulus becomes
\begin{align}
&\frac{i}{2\pi}\int_{\varepsilon<|\bm k|<R}\Tr\bigl(P[P_x,P_y]\bigr)\,d^2k
\notag\\
&\qquad=\frac{1}{2\pi}\int_0^{2\pi}\!\!\int_\varepsilon^R
i\Tr\bigl(P[P_r,P_\theta]\bigr)\,dr\,d\theta .
\label{eq:annulus}
\end{align}

\subsection{The bridge identity}

\begin{lemma}
\label{lem:bridge}
If $P$ is a differentiable family of orthogonal projectors with $P_\theta=i[D,P]$, then $i\Tr\bigl(P[P_r,P_\theta]\bigr)=-\partial_r\Tr(DP)$.
\end{lemma}

\begin{proof}
\emph{Step 1.} Differentiating $P^2=P$ gives
\begin{equation}
P_rP+PP_r=P_r.
\label{eq:Pr}
\end{equation}
Multiplying Eq.~\eqref{eq:Pr} by $P$ on both sides gives $2PP_rP=PP_rP$, so $PP_rP=0$.

\emph{Step 2.} Insert $P_\theta=i[D,P]$: $i\Tr\bigl(P[P_r,P_\theta]\bigr)=-\Tr\bigl(P[P_r,[D,P]]\bigr)$. Expand the commutators into four products and use $PP_rP=0$ and $P^2=P$:
\begin{align}
\Tr(PP_rDP)&=\Tr(PP_rD), & \Tr(PP_rPD)&=0,\notag\\
\Tr(PDPP_r)&=0, & \Tr(PPDP_r)&=\Tr(PDP_r).
\end{align}
The first entry moves the final $P$ to the front by cyclicity, and the third uses $\Tr(PDPP_r)=\Tr(DPP_rP)$. Hence
\begin{align}
\Tr\bigl(P[P_r,[D,P]]\bigr)&=\Tr(PP_rD)+\Tr(PDP_r)\notag\\
&=\Tr\bigl(D(PP_r+P_rP)\bigr)=\Tr(DP_r),
\end{align}
by cyclicity and Eq.~\eqref{eq:Pr}.

\emph{Step 3.} $D$ does not depend on $r$, so $\Tr(DP_r)=\partial_r\Tr(DP)$.
\end{proof}

No eigenvectors, and hence no phase choices, appear in this proof. By Lemma~\ref{lem:bridge} the right side of Eq.~\eqref{eq:annulus} equals $-\int_\varepsilon^R W_B'(r)\,dr$, so the curvature over the annulus is exactly $W_B(\varepsilon)-W_B(R)$.

\subsection{The two ends}

\emph{Small $r$.} $P$ is smooth at $\bm k=0$, so the integrand $\Tr(P[P_x,P_y])$ is bounded near the origin and the disk $|\bm k|<\varepsilon$ contributes at most a constant times $\varepsilon^2$. With $W_B(\varepsilon)\to W_B(0)$, the curvature over $|\bm k|<R$ equals $W_B(0)-W_B(R)$.

\emph{Large $r$.} For $r>0$, $H_B(0,r)=r\bigl(\Sigma_1+tV_B\bigr)$ with $t=1/r$, and a positive factor does not change spectral projectors. Let $Q(t)$ be the projector onto the $m$ lowest eigenvalues of $\Sigma_1+tV_B$, so $P(0,r)=Q(1/r)$. For $|t|\,\|V_B\|<\tfrac12$, Weyl's inequality keeps those eigenvalues within $\tfrac12$ of $-1$ and the others within $\tfrac12$ of $+1$. The circle of radius 1 around $-1$ therefore serves as $\Gamma$ for all such $t$, and $Q$ is smooth in $t$, including at $t=0$. Hence $\|P_r\|=r^{-2}\|Q'(1/r)\|\le Cr^{-2}$ at angle zero, and by Eq.~\eqref{eq:P-covariance} at every angle, while $\|P_\theta\|\le2\|D\|=2m$. The integrand of Eq.~\eqref{eq:annulus} is therefore at most a constant times $r^{-2}$, the integral over $r>R$ tends to zero, and
\begin{equation}
I_B=W_B(0)-W_B(\infty),\qquad W_B(\infty)=\Tr\bigl(DQ(0)\bigr).
\end{equation}
$Q(0)$ projects onto the $-1$ eigenspace of $\Sigma_1=I_m\otimes\sigma_1$, so $Q(0)=I_m\otimes\tfrac12(I_2-\sigma_1)$, whose diagonal entries all equal $\tfrac12$. With $D=\bigoplus_j\diag(j-1,j)$,
\begin{equation}
W_B(\infty)=\sum_{j=1}^m\frac{(j-1)+j}{2}=\frac{m^2}{2}.
\end{equation}
$Q(0)$ does not involve $V_B$, so the two bulks have the same limit $P(\theta,\infty)=I_m\otimes\tfrac12(I_2-\cos\theta\,\sigma_1-\sin\theta\,\sigma_2)$ in every direction. This is where ellipticity enters: both bulks share the principal symbol $k_x\Sigma_1+k_y\Sigma_2$.

\subsection{Zero momentum}

At $\bm k=0$, $H_B(\bm0)=V_B$ is block diagonal, with blocks $\operatorname{span}\{A_1\}$, $\operatorname{span}\{B_j,A_{j+1}\}$ for $1\le j\le m-1$, and $\operatorname{span}\{B_m\}$. On these blocks $D$ equals $0$, $jI_2$, and $m$. Choose an orthonormal eigenbasis of $V_B$ made of eigenvectors of the individual blocks; each basis vector is then also an eigenvector of $D$, with eigenvalue equal to its angular weight. By the direct gap, the $m$ lowest eigenvalues are separated from the rest, so $P_B(\bm0)$ is the sum of the projectors onto the corresponding $m$ basis vectors, whatever basis is chosen inside a degenerate eigenspace. Hence
\begin{equation}
W_B(0)=\Tr\bigl(DP_B(\bm0)\bigr)=\sum_{\text{occupied }a}w_a,
\end{equation}
which proves Theorem~\ref{thm:weights}. Each $W_B(0)$ is an integer, so $I_B\in\mathbb Z-m^2/2$ and $\mathfrak C=W_N(0)-W_S(0)$ is an integer, with no further argument needed. In the other valley, covariance is generated by $-D$ and every $I_B$ changes sign.

\section{Equal-coupling lemmas}
\label{app:equal}

\begin{lemma}[Family ordering]
\label{lem:order}
For nondecreasing $\bm u$ and every common $\gamma>0$, $E^\pm_j\le E^\pm_{j+1}$, with equality exactly when $u_j=u_{j+1}=u_{j+2}$.
\end{lemma}

\begin{proof}
With $x=u_{j+1}-u_j\ge0$, $y=u_{j+2}-u_{j+1}\ge0$, and $S=\sqrt{x^2+4\gamma^2}+\sqrt{y^2+4\gamma^2}$, rationalizing the difference of square roots in Eq.~\eqref{eq:pair} gives
\begin{equation}
E^\pm_{j+1}-E^\pm_j=\frac{x+y}{2}\Bigl(1\pm\frac{y-x}{S}\Bigr).
\end{equation}
Since $S>x+y\ge|y-x|$, both factors in parentheses are positive.
\end{proof}

\begin{lemma}[Prefix structure]
\label{lem:prefix}
For nondecreasing $\bm u$ at equal coupling with a direct gap, the occupied levels are $u_1$, $E^-_1,\ldots,E^-_q$, and $E^+_1,\ldots,E^+_p$, with $p+q=m-1$ and $0\le p\le p_{\max}$.
\end{lemma}

\begin{proof}
The first part of the proof of Theorem~\ref{thm:configurations} shows that $u_1$ is occupied and $u_m$ is unoccupied. Lemma~\ref{lem:order} makes the occupied upper levels and the occupied lower levels each a prefix, and tied levels cannot straddle the gap. Admissibility of $(\{1,\ldots,p\},\{q+1,\ldots,m-1\})$ requires $p+2\le q+1$, that is, $p<q$.
\end{proof}

\begin{lemma}[Linear realizations]
\label{lem:linear}
Let $m\ge2$, $\gamma>0$ and $0\le p\le p_{\max}$. There is $s>0$ such that the bulk with $u_j=s\bigl(j-\frac{m+1}2\bigr)$ and $\gamma_j=\gamma$ has a direct gap and $U=\{1,\ldots,p\}$, $C=\{m-p,\ldots,m-1\}$.
\end{lemma}

\begin{proof}
Every pair has the same half-splitting; put $h=h_j/s=\sqrt{1/4+\gamma^2/s^2}$. As $s$ runs over $(0,\infty)$, $h$ runs over $(\tfrac12,\infty)$, and $s=\gamma/\sqrt{h^2-1/4}$ inverts the map, so it suffices to choose $h$. Eq.~\eqref{eq:pair} gives
\begin{equation}
E^\pm_j=s\Bigl(j-\frac m2\pm h\Bigr),\qquad u_m=-u_1=s\,\frac{m-1}2>0 .
\end{equation}
Hence $E^+_j<0\iff j<\frac m2-h$ and $E^-_j<0\iff j<\frac m2+h$. Choose
\begin{equation}
\max\Bigl\{\tfrac12,\ \tfrac m2-p-1\Bigr\}<h<\tfrac m2-p .
\label{eq:h-window}
\end{equation}
The interval is nonempty because $p\le p_{\max}<\frac{m-1}2$ gives $\frac m2-p>\frac12$. Then $\frac m2-h\in(p,p+1)$ and $\frac m2+h\in(m-p-1,m-p)$, so the negative levels are exactly $u_1$, $E^-_1,\ldots,E^-_{m-1-p}$ and $E^+_1,\ldots,E^+_p$. There are $1+(m-1-p)+p=m$ of them, and no level is zero because $\frac m2\pm h$ is not an integer. The zero-momentum gap is therefore open, and Lemma~\ref{lem:zero-gap} gives a direct gap. The occupied upper levels are $E^+_1,\ldots,E^+_p$ and the unoccupied lower levels are $E^-_{m-p},\ldots,E^-_{m-1}$.
\end{proof}

\begin{lemma}[Reflection symmetry]
\label{lem:J}
Let $R$ reverse the $2m$ components and $Z=\diag(1,-1,1,\ldots,-1)$. If $u_{m+1-j}=-u_j$ and $\gamma_{m-j}=\gamma_j$, then $J=ZR$ satisfies $JH_B(r,0)J^\dagger=-H_B(r,0)$ for all $r$, and by covariance at every momentum. The ordered spectrum obeys $e_{2m+1-\ell}=-e_\ell$, and a direct gap $\delta_B$ implies that the spectrum avoids $(-\delta_B/2,\delta_B/2)$.
\end{lemma}

\begin{proof}
$R$ maps the diagonal to its reverse, which is its negative, and preserves the palindromic off-diagonal sequence $(r,\gamma_1,r,\ldots,\gamma_{m-1},r)$. $Z$ preserves the diagonal and negates every nearest-neighbor off-diagonal entry.
\end{proof}

\section{Gap lower bounds and asymmetric bulks}
\label{app:cert}

Because $\partial_rH_B(r,0)=\Sigma_1$ has norm one, each ordered eigenvalue $e_\ell(r)$ is 1-Lipschitz. Sampling $r\in[0,\|V_B\|+5E_0]$ with spacing $\delta r$ and widening the sampled extrema by $\delta r/2$ bounds $\sup e_m$ from above and $\inf e_{m+1}$ from below on that interval. Beyond it, the bounds in the proof of Lemma~\ref{lem:zero-gap} keep both bands outside $(-4E_0,4E_0)$. With $\delta r=2\times10^{-4}E_0$ this gives the windows of Table~\ref{tab:examples}, exact up to floating-point rounding in the $16\times16$ eigenvalue computations. Table~\ref{tab:asym} lists the asymmetric bulks that complete $\cD^{\mathrm{glob}}_m$, with the same lower bounds. We found them by maximizing the global gap at fixed $(U,C)$ and report them to three decimals, the precision at which we computed the bounds.

\begin{table*}
\caption{Ascending bulks with a global gap, for values of $\Delta$ absent from the reflection-symmetric family. Window: interval avoided by the spectrum, from the lower bound above. Energies in units of $E_0$.}
\label{tab:asym}
\begin{ruledtabular}
\begin{tabular}{cclll}
$m$ & $\Delta$ & $(U,C)$ & $\bm u$ \;;\; $\bm\gamma$ & window\\
\hline
5 & 2 & $(\{1\},\{3\})$ & $(-3,-3,2,2,2)$ ; $(1.5,2,1.5,2.5)$ & $(-0.442,0.095)$\\
6 & 3 & $(\{1\},\{4\})$ & $(-2.525,-2.525,0.934,1.366,1.368,1.381)$ ; $(0.982,1.621,1.548,1.210,1.861)$ & $(-0.413,-0.018)$\\
7 & 2 & $(\{1\},\{3\})$ & $(-2.078,-2.069,0.823,0.823,0.825,0.825,0.851)$ ; $(1.133,1.977,0.804,2.061,1.397,1.329)$ & $(-0.340,0.008)$\\
7 & 4 & $(\{1\},\{5\})$ & $(-1.550,-1.507,-0.542,0.528,1.005,1.005,1.060)$ ; $(0.610,1.006,1.505,1.340,0.746,1.216)$ & $(-0.183,0.056)$\\
\end{tabular}
\end{ruledtabular}
\end{table*}

\section{Numerical details}
\label{app:numerics}

Re-orthonormalization and double precision are essential. An earlier implementation propagated unorthonormalized frames in single precision with an adaptive Dormand--Prince integrator (relative tolerance $10^{-7}$, at the single-precision machine epsilon) and orthonormalized only at the matching point. It reproduces the three-layer wall, with crossing positions agreeing to $3\times10^{-3}$. For wall vi, however, it reports 68 near-intersections with matching score above 1.9, where the correct count is 16. The spurious ones cluster at $|k_y|\approx3$--$5.7$, where the growth rates are most disparate and the frames collapse onto the dominant mode. Its default window $|k_y|\le1.5$ would also miss most of the genuine crossings. For walls with small energy scales, such as wall vii, the same implementation is accurate: its ten crossings agree in position with our double-precision calculation for the same profile to within $\pm0.02$. There the decay rates are close together and the frames do not collapse.

\begin{figure}
\includegraphics[width=\columnwidth]{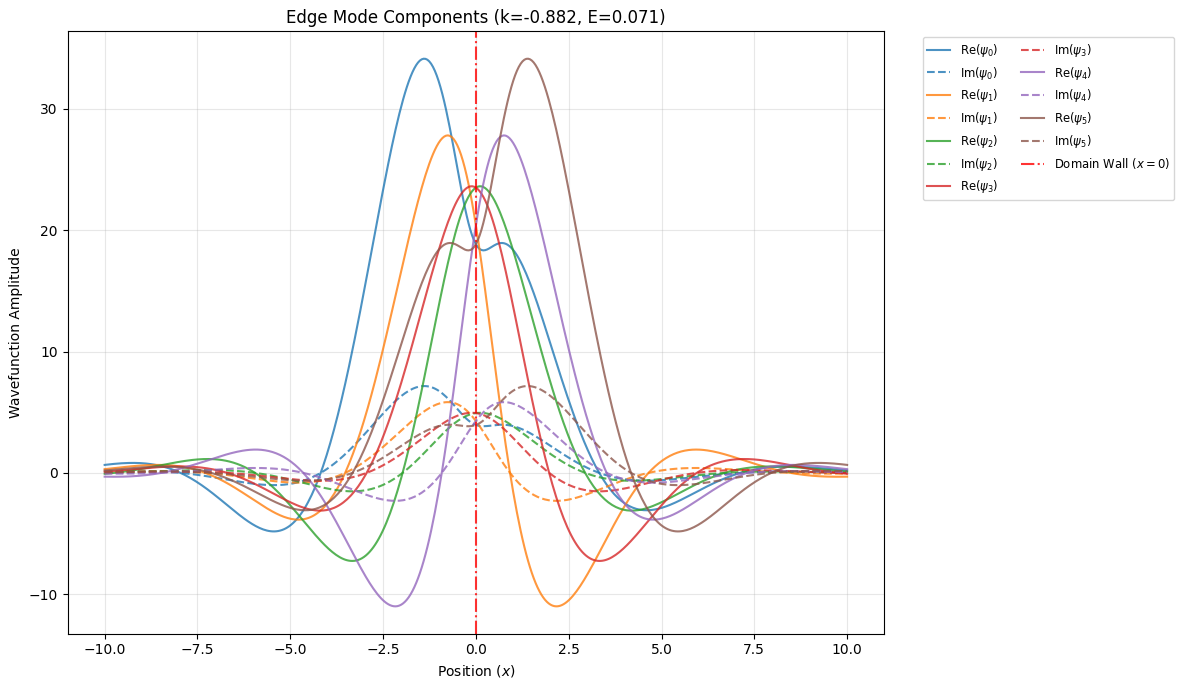}
\caption{Components $\psi_0,\ldots,\psi_5$ (implementation basis, $A_1,B_1,\ldots,B_3$; arbitrary normalization) of an interface eigenfunction of the three-layer wall of Appendix~\ref{app:numerics}, at implementation momentum $k=-0.882E_0/\hbar v$ and $E=0.0714E_0$. The decomposition into real and imaginary parts depends on the basis; the localization at the wall does not.}
\label{fig:edgemode}
\end{figure}

Figure~\ref{fig:edgemode} shows an interface eigenfunction from that implementation for the three-layer wall $u_j(y)=\tanh(2y)(-1,0,1)_jE_0$, $\gamma_j=E_0$, which has $\mathfrak C=-3$. The plotted point lies within $2\times10^{-5}E_0$ of an eigenvalue, and the mode is confined to a few $\hbar v/E_0$ of the wall.

\end{document}